\documentclass[12pt]{article}

\usepackage{epsfig,subfigure,graphicx,amsmath,amsfonts}
\newfont{\bbb}{msbm10 scaled 500}

\newfont{\bb}{msbm10 scaled 1100}
\newcommand{\CC}{\mbox{\bb C}}
\newcommand{\RR}{\mbox{\bb R}}
\newcommand{\ZZ}{\mbox{\bb Z}}

\newcommand{\cv}{{\bf c}}

\newcommand{\rv}{{\bf r}}

\newcommand{\tv}{{\bf t}}
\newcommand{\uv}{{\bf u}}
\newcommand{\wv}{{\bf w}}
\newcommand{\vv}{{\bf v}}
\newcommand{\xv}{{\bf x}}
\newcommand{\yv}{{\bf y}}
\newcommand{\zv}{{\bf z}}
\newcommand{\zerov}{{\bf 0}}

\newcommand{\Cm}{{\bf C}}

\newcommand{\Gm}{{\bf G}}
\newcommand{\Hm}{{\bf H}}
\newcommand{\Id}{{\bf I}}

\newcommand{\Mm}{{\bf M}}

\newcommand{\Pm}{{\bf P}}
\newcommand{\Qm}{{\bf Q}}
\newcommand{\Rm}{{\bf R}}
\newcommand{\Sm}{{\bf S}}
\newcommand{\Tm}{{\bf T}}

\newcommand{\Ac}{{\cal A}}

\newcommand{\Cc}{{\cal C}}

\newcommand{\Nc}{{\cal N}}

\newcommand{\Rc}{{\cal R}}

\newcommand{\Uc}{{\cal U}}

\newcommand{\lambdav}{\hbox{\boldmath$\lambda$}}

\newcommand{\Lambdam}{\hbox{\boldmath$\Lambda$}}

\newcommand{\Sigmam}{\hbox{\boldmath$\Sigma$}}

\newcommand{\Thetam}{\hbox{\boldmath$\Theta$}}

\newcommand{\sign}{{\hbox{sign}}}
\renewcommand{\arg}{{\hbox{arg}}}

\renewcommand{\Re}{{\rm Re}}
\renewcommand{\Im}{{\rm Im}}

\newcommand{\transp}{{^T}}
\newcommand{\sparse}{{\textsf{S}}}

\newcommand{\defines}{{\,\,\stackrel{\scriptscriptstyle \bigtriangleup}{=}\,\,}}

\newcommand{\be}{\begin{equation}}
\newcommand{\ee}{\end{equation}}
\newcommand{\bear}{\begin{eqnarray}}
\newcommand{\eear}{\end{eqnarray}}
\newcommand{\mb}{\mathbf }
\newcommand{\mc}{\mathcal}
\newcommand{\bmx}{\begin{pmatrix}}
\newcommand{\emx}{\end{pmatrix}}

\newtheorem{theorem}{Theorem}
\newtheorem{proposition}{Proposition}
\newenvironment{proof}{{\sl Proof \/}:\ \ }

\begin{document}

\title{A Unified Framework for Tree Search Decoding: Rediscovering the
  Sequential Decoder}
\author{A.~D.~Murugan, H.~El~Gamal, M. O. Damen and
  G. Caire \thanks{Arul~D.~Murugan
and Hesham~El~Gamal are with the ECE Department at the Ohio State
University. Mohamed~Oussama~Damen is with the ECE Department at
the the University of Waterloo. Giuseppe~Caire is with The Mobile
Communication group at Eurecom Institute. The work of
Arul~.D.~Murugan and Hesham~El~Gamal was supported partly by NSF
CAREER grant~0346887 and a gift from Texas Instruments. }}
\maketitle

%-----------------------------------------------------------------------------------
%-----------------------------------------------------------------------------------

\begin{abstract}
%%% last edit: Giuseppe, 04/26/05
We consider receiver design for coded transmission over linear
Gaussian channels. We restrict ourselves to the class of lattice
codes and formulate the joint detection and decoding problem as a
closest lattice point search (CLPS). Here, a tree search framework
for solving the CLPS is adopted. In our framework, the CLPS
algorithm decomposes into the preprocessing and tree search
stages. The role of the preprocessing stage is to expose the tree
structure in a form {\em matched} to the search stage. We
argue that the minimum mean square error decision feedback
(MMSE-DFE) frontend is instrumental for solving the joint
detection and decoding problem in a single search stage. It is
further shown that MMSE-DFE filtering allows for using lattice
reduction methods to reduce complexity, at the expense of a
marginal performance loss, and solving under-determined linear
systems. For the search stage, we present a generic method, based
on the branch and bound (BB) algorithm, and show that it
encompasses all existing sphere decoders as special cases. The
proposed generic algorithm further allows for an interesting
classification of tree search decoders, sheds more light on the
structural properties of all known sphere decoders, and inspires
the design of more efficient decoders. In particular, an efficient
decoding algorithm that resembles the well known Fano sequential
decoder is identified. The excellent performance-complexity
tradeoff achieved by the proposed MMSE-Fano decoder is established
via simulation results and analytical arguments in several MIMO
and ISI scenarios.
\end{abstract}

%-----------------------------------------------------------------------------------
%-----------------------------------------------------------------------------------

\section{Introduction} \label{intro}
%%% last edit: Giuseppe, 04/26/05

Recent years have witnessed a growing interest in the closest
lattice point search (CLPS) problem. This interest was primarily
sparked by the connection between CLPS and maximum likelihood (ML)
decoding in multiple-input multiple-output (MIMO) channels
\cite{damen0}. On the positive side, MIMO channels offer
significant advantages in terms of increased throughput and
reliability. The price entailed by these gains, however, is a more
challenging decoding task for the receiver. For example, naive
implementations of the ML decoder have complexity that grows
exponentially with the number of transmit antennas. This
observation inspired several approaches for sub-optimal decoding
that offer different performance-complexity tradeoffs (e.g.,
\cite{babai,foschini}).

Reduced complexity decoders are typically obtained by exploiting
the codebook structure. The scenario considered in our work is no
exception. In principle, the decoders considered here exploit the
underlying lattice structure of the received signal to cast the decoding
problem as a CLPS. Some variants of such decoders are known in the literature
as {\em sphere decoders} (e.g.,
\cite{damen1,hassibi,agrell,radhika1,vit-bou}). These decoders
typically exploit number-theoretic ideas to efficiently span the
space of allowed codewords (e.g., \cite{pohst,schn-euch}). The
complexity of such decoders were shown, via simulation and
numerical analysis, to be significantly smaller than the naive ML
decoder in many scenarios of practical interest (e.g.,
\cite{damen1,hassibi}). The complexity of the state of the art
sphere decoder, however, remains prohibitive for problems
characterized by a large dimensionality \cite{otter}. This
observation is one of the main motivations for our work.

The overriding goal of our work is to establish a general
framework for the design and analysis of tree search algorithms
for joint detection and decoding. Towards this goal, we first
divide the decoding task into two interrelated stages; namely, 1)
preprocessing and 2) tree search. The preprocessing stage is
primarily concerned with exposing the underlying tree structure
from the noisy received signal. Here, we discuss the integral
roles of minimum mean square error decision feedback (MMSE-DFE)
filtering, lattice reduction techniques, and relaxing the boundary
control (i.e., lattice decoding) in tree search decoding. We then
proceed to the search stage where a general framework based on the
branch and bound (BB) algorithm is presented. This framework
establishes, rigorously, the equivalence in terms of performance
and complexity between different sphere and sequential decoders.
We further use the proposed framework to classify the different
search algorithms and identify their advantages/disadvantages. The
MMSE-Fano decoder emerges as a special case of our general
framework that enjoys a favorable performance-complexity tradeoff.
We establish the superiority of the proposed decoder via numerical
results and analytical arguments in several relevant scenarios
corresponding to coded as well as uncoded transmission over MIMO
and inter-symbol-interference (ISI) channels. More specifically,
in our simulation experiments, we apply the tree search decoding
framework to uncoded V-BLAST \cite{vblast}, linear dispersion
space-time codes \cite{LD}, algebraic space-time codes
\cite{RHHG,LK,LFT}, and trellis codes over ISI channels \cite{isi}. In all
these cases, our results show that the MMSE-Fano decoder achieves near-ML
performance with a much smaller complexity.

The rest of the paper is organized as follows.
Section~\ref{system} introduces our system model and notation. In
Section~\ref{prep}, we consider the design of the preprocessing
stage and discuss the interplay between this stage and the tree
search stage. In Section~\ref{search_algos}, we present a general
framework for designing tree search decoders based on the branch
and bound (BB) algorithm. In Section~\ref{analysis}, we establish
the superior performance-complexity tradeoff achieved by the
proposed MMSE-Fano decoder, using analytical arguments and
numerical results, in several interesting scenarios. Finally, we offer some
concluding remarks in Section~\ref{conclusions}

%-----------------------------------------------------------------------------------
%-----------------------------------------------------------------------------------

\section{System Model} \label{system}
%%% last edit: Giuseppe, 04/26/05

We consider the transmission of lattice codes over linear channels
with white Gaussian additive noise (AWGN). The importance of this
problem stems from the fact that several very relevant
applications arising in digital communications fall in this class,
as it will be illustrated by some examples at the end of this
section. Let $\Lambda \subseteq\RR^m$ be an $m$-dimensional
lattice, i.e., the set of points
\begin{equation} \label{lattice}
\Lambda = \{ \lambdav = \Gm \xv \; : \; \xv \in \ZZ^m \}
\end{equation}
where $\Gm \in \RR^{m\times m}$ is the lattice generator matrix.
Let $\vv \in \RR^m$ be a vector and $\Rc$ a measurable region in
$\RR^m$. A lattice code $\Cc(\Lambda,\vv,\Rc)$ is defined
\cite{forney-coset1,forney-coset2,hesham1} as the set of points of
the lattice translate $\Lambda + \vv$ inside the {\em shaping
  region} $\Rc$, i.e.,
\begin{equation} \label{code}
\Cc(\Lambda,\vv,\Rc) = \{\Lambda + \vv\} \cap \Rc.
\end{equation}
Without loss of generality, we can also see $\Cc(\Lambda,\vv,\Rc)$
as the set of points $\cv + \vv$, such that the {\em codewords}
$\cv$ are given by
\begin{equation} \label{code1}
\cv = \Gm \xv, \;\;\; \mbox{for} \;\; \xv \in \Uc
\end{equation}
where $\Uc \subset \ZZ^m$ is the code {\em information set}.

The linear additive noise channel is described, in general, by the
input-output relation
\begin{equation} \label{sys0}
\mathbf{r = H (c+v) + z}
\end{equation}
where $\rv \in \RR^n$ denotes the received signal vector, $\zv
\sim \Nc(\zerov,\Id)$ is the AWGN vector, and $\Hm \in {\mathbb
R}^{n\times m}$ is a matrix that defines the channel linear
mapping between the input and the output.

Consider the following communication problem: a vector of
information symbols $\xv$ is generated with uniform probability
over $\Uc$, the corresponding codeword $\cv = \Gm\xv$ is produced
by the encoder and the signal $\cv + \vv$ is transmitted over the
channel (\ref{sys0}). Assuming $\Hm$ and $\vv$ known to the
receiver, the ML decoding rule is given by
\begin{equation}
\label{sys0a} \hat{\mb x} = \textrm{arg} \min_{{\mb x}\in
{\mathcal U}} |{\mathbf r}-{\Hm}{\vv}-{\mb H}\Gm {\mb x}|^2
\end{equation}
%where we define the translated received signal
%\begin{equation} \label{y}
%\yv = \rv - \Hm\vv.
%\end{equation}
%Clearly, $\yv$ is a sufficient statistics for the detection of
%$\xv$ and from now on it shall be identified as the {\em channel
%output}. 
The constraint ${\mathcal U}\subset {\mathbb
  Z}^m$ implies that
the optimization problem in (\ref{sys0a}) can be viewed as a {\em
  constrained} version of the CLPS with lattice
generator matrix given by ${\mathbf H}\Gm$ and constraint set
$\Uc$.

A few remarkable examples of the above framework are:

\begin{enumerate}

\item{\it MIMO flat fading channels:} One of the simplest and most
widely studied examples is a MIMO V-BLAST system with squared QAM
modulation, $M$ transmit and $N$ receive antennas, operating over
a flat Rayleigh fading channel. The baseband complex received
signal\footnote{We use the superscript $^c$ to denote complex
variables.} in this case can be expressed as
\begin{equation} \label{vblast1}
{\mb r}^c=\sqrt{\frac{\rho}{M}}{\mb H}^c\mb{c}^c+\mb{z}^c
\end{equation}
where the complex channel matrix ${\mb H}^c\in {\mathbb
C}^{N\times M}$ is composed of i.i.d elements $h_{i,j}^c\sim
{\mathcal N}_{\mathcal C}(0,1)$, the input complex signal ${\mb
c}^c$ has components ${\mb c}_i^c$ chosen from a unit-energy
$Q^2$-QAM constellation, the noise has i.i.d. components $z^c_i
\sim \Nc_{\Cc}(0,1)$ and $\rho$ denotes the signal to noise ratio
(SNR) observed at any receive antenna. The system model in
(\ref{vblast1}) can be expressed in the form of (\ref{sys0}) by
appropriate scaling and by separating the real and imaginary parts
using the vector and the matrix transformations defined by
\[ \uv^c \mapsto \uv = [\Re\{\uv^c\}^\transp, \Im\{\uv^c\}^\transp]^\transp, \]
\[
{\mb M}^c \mapsto \Mm = \left [ \begin{array}{cc}
\Re\{\Mm^c\} & - \Im\{\Mm^c\} \\
\Im\{\Mm^c\} & \Re\{\Mm^c\} \end{array} \right ].
\]
The resulting real model is given by (\ref{sys0}) where $n = 2N$,
$m = 2M$ and the constraint set is given by $\Uc = \ZZ_Q^m$, with
$\ZZ_Q = \{0,\ldots,Q-1\}$ denoting the set of integers residues
modulo $Q$.

In the case of V-BLAST, the lattice code generator matrix $\Gm =
\kappa\Id$, where $\kappa$ is a normalizing constant, function of
$Q$, that makes the (complex) transmitted signal of unit energy
per symbol. This formulation extends naturally to MIMO channels
with more general lattice coded inputs \cite{hesham1}. In general,
a space-time code of block length $T$ is defined by a set of
matrices $\Cm^c = [\cv^c_1,\ldots,\cv^c_T]$ in $\CC^{M\times T}$.
The columns of the codeword $\Cm^c$ are transmited in parallel on
the $M$ transmit antennas in $T$ channel uses. The received signal
is given by the sequence of vectors
\begin{equation} \label{vblast2}
{\mb r}^c_t =\sqrt{\frac{\rho}{M}}{\mb H}^c\mb{c}^c_t +\mb{z}^c_t,
\;\;\; t = 1,\ldots,T
\end{equation}
Lattice space-time codes are obtained by taking a lattice code
$\Cc(\Lambda,\vv,\Rc)$ in $\RR^{2MT}$, and mapping each codeword
$\cv$ into a complex matrix $\Cm^c$ according to some {\em linear}
one-to-one mapping $\RR^{2MT} \rightarrow \CC^{M \times T}$. It is
easy to see that a lattice-coded MIMO system can be again
expressed by (\ref{sys0}) where the channel matrix $\Hm$ is
proportional (through an appropriate scaling factor) to the
block-diagonal matrix
\begin{equation}
\label{sys2a} \Id_T \otimes \left [ \begin{array}{cc}
\Re\{\mb{H}^c\} & -\Im\{\mb{H}^c\} \\
\Im\{\mb{H}^c\} & \Re\{\mb{H}^c\}
\end{array}\right]
\end{equation}
In this case, we have $n = 2NT$ and $m = 2MT$. It is interesting
to notice that for a wide class of {\em linear dispersion} (LD)
codes
\cite{LD,tast,tast-constellations,golden-code,division-algebra},
the information set $\Uc$ is still given by $\ZZ_Q^m$, as in the
simple V-BLAST case, although the generator matrix $\Gm$ is
generally not proportional to $\Id$. For other classes of lattice
codes \cite{hesham1}, with more involved shaping regions $\Rc$,
the information set $\Uc$ does not take on the simple form of an
``hypercube''. For example, consider $\Lambda$ obtained by
construction A \cite{splug}, i.e., $\Lambda = C + Q\ZZ^m$, where
$C \subseteq \ZZ_Q^m$ is a linear code over $\ZZ_Q$ with generator
matrix in systematic form $[\Id, \Pm^\transp]^\transp$. A
generator matrix of $\Lambda$ is given by \cite{splug}
\begin{equation} \label{gmatrix}
\Gm = \left [ \begin{array}{cc}
\Id & \zerov \\
\Pm & Q \Id \end{array} \right ].
\end{equation}
Typically, the shaping region $\Rc$ of the lattice code
$\Cc(\Lambda,\vv,\Rc)$ can be an $m$-dimensional sphere, the
fundamental Voronoi region of a sublattice $\Lambda' \subset
\Lambda$, or the $m$-dimensional hypercube. In all these cases,
the information set $\Uc$ may be difficult to describe.

\item {\it ISI Channels}: For simplicity, we consider a baseband
{\em real} single-input single-output  (SISO)
inter-symbol-interference (ISI) channel with the input and output
sequences related by \[{r_i = \sum_{\ell
    = 0}^{L} h_\ell c_{i-\ell} + z_i},\] where
$(h_0,\ldots,h_L)$ denotes the discrete-time channel impulse
response, assumed of finite length $L+1$. The extension to the
complex baseband model is immediate. Assuming that the transmitted
signal is padded by $L$ zeroes, the channel can be written in the
form (\ref{sys0}) where the channel matrix takes on the tall
banded Toeplitz form
\[
\mathbf{H} = \left [
\begin{array}{cccc}
h_0 &  &  &                 \\
h_1 & h_0 &  &              \\
\vdots & \ddots & \ddots &  \\
h_L & \ddots & \ddots & h_0 \\
   & h_L & \ddots & h_1     \\
   &    &  \ddots & \vdots  \\
   &   &          & h_L     \\
\end{array}\right ].
\]
A wide family of trellis codes obtained as coset-codes
\cite{forney-coset1,forney-coset2}, including binary linear codes,
can be formulated as lattice codes where $\Lambda$ is a
Construction A lattice and  the shaping region $\Rc$ is chosen
appropriately. In particular, coded modulation schemes based on
the $Q$-PAM constellation obtained by mapping group codes over
$\ZZ_Q$ onto the $Q$-PAM constellation can be seen as lattice
codes with hypercubic shaping $\Rc$. The important case of binary
convolutional codes falls in this class for $Q = 2$. Again, the
information set $\Uc$ corresponding to $\Rc$ may, in general, be
very complicated.

\end{enumerate}

%-----------------------------------------------------------------------------------
%-----------------------------------------------------------------------------------

\section{The Preprocessing Stage} \label{prep}
%%% last edit: Giuseppe, 04/26/05

In our framework, we  divide the CLPS into two stages; namely, 1)
preprocessing and 2) tree search.  The complexity and performance
of CLPS algorithms depend critically on the efficiency of the
preprocessing stage. Loosely, the goal of preprocessing is to
transform the original constrained CLPS problem, described by the
lattice generator matrix $\Hm \Gm$  and by the constraint set
${\Uc}$, into a form which is {\em friendly} to the search
algorithm used in the subsequent stage. In the following, we
discuss the different tasks performed in the preprocessing stage.
In general, a {\em friendly} tree structure can be exposed through
three steps: left preprocessing, right preprocessing, and forming
the tree.

Some options for these three steps are illustrated in the
following subsections. However, before entering the algorithmic
details, it is worthwhile to point out some general
considerations. The classical sphere decoding approach to
the solution of the original constrained  CLPS problem
(\ref{sys0a}) consists of applying QR decomposition on the
combined channel and code matrix, i.e., letting $\Hm\Gm = \Qm\Rm$
where $\Qm \in \RR^{n \times m}$ has orthonormal columns and $\Rm
\in \RR^{m
  \times m}$ is upper triangular.
Using the fact that $\Qm^\transp$ is an isometry with respect to
the Euclidean distance, (\ref{sys0a}) can be written as
\begin{equation}
\label{sys0a-qr} \hat{\mb x} = \textrm{arg} \min_{{\mb x}\in
{\mathcal U}} \Big |{\mathbf y}'-\Rm \xv \Big |^2
\end{equation}
where $\yv' = \Qm^\transp (\rv-\Hm\vv)$. If rank$(\Hm\Gm) = m$, $\Rm$ has
non-zero diagonal elements and its triangular form can be
exploited to search for all the points $\xv \in \Uc$ such that
$\Rm\xv$ is in a sphere of a given search radius centered in
$\yv'$. If the sphere is non-empty, the ML solution is guaranteed
to be found inside the sphere, otherwise, the search radius is
increased and the search is restarted. Different variations on
this main theme have been proposed in the literature, and will be
reviewed in Section \ref{search_algos} as  special cases of a
general BB algorithm.  Nevertheless, it is useful to point out
here the two main sources of inefficiency of the above approach:
1) It does not apply to  the case rank$(\Hm\Gm) < m$ and, even
when rank$(\Hm\Gm) = m$ but $\Hm\Gm$ is ill-conditioned, the
spread (or dynamic range) of the diagonal elements  of $\Rm$ is
large. This entails large complexity of the tree search
\cite{damen2}. Intuitively, when $\Hm\Gm$ is ill-conditioned, the
lattice generated by $\Hm\Gm$ has a very skewed fundamental cell
such that there are directions in which it is very difficult to
distinguish the points $\{\Hm\Gm\xv : \xv \in \Uc\}$; 2) Enforcing
the condition $\xv \in \Uc$, can be very difficult because a
lattice code $\Cc(\Lambda,\vv,\Rc)$ with non-trivial shaping
region $\Rc$ might have an information set $\Uc$ with a
complicated shape. Hence, just checking the condition $\xv \in
\Uc$ during the search may entail a significant complexity.

Left preprocessing can be seen as an effort to tackle the first
problem: it modifies the channel matrix and the noise vector such
that the resulting CLPS problem is non-equivalent to ML
(therefore, it is suboptimal), but it has a much better
conditioned ``channel'' matrix. The second problem can be tackled
by relaxing the constraint set $\Uc$ to the whole $\ZZ^m$, i.e.,
searching over the whole lattice $\Lambda$ instead of only the
lattice code $\Cc$ (or lattice decoding). In general, lattice
decoding is another source of suboptimality. Nevertheless, once
the boundary region is removed, we have the freedom of choosing
the lattice basis which is more convenient for the search
algorithm. This change of lattice basis is accomplished by right
preprocessing. Finally, the tree structure is obtained by
factorizing the resulting combined channel-lattice matrix in upper
triangular form, as in classical sphere decoding. Overall, left
and right preprocessing combined with lattice decoding are a way
to reduce complexity at the expense of optimality. Fortunately, it
turns out that an appropriate combination of these elements yields
very significant saving in complexity with very small degradation
with respect to the ML performance. Thus, it yields a very
attractive decoding solution. While the outstanding performance of
appropriate preprocessing and lattice decoding can be motivated
via  rigorous information theoretic arguments
\cite{hesham1,erez-zamir,erez-zamir-litsyn}, here we are more
concerned with the algorithmic aspects of the decoder and we shall
give some heuristic motivation based on ``signal-processing''
arguments.

Finally, we note that the notion of complexity adopted in this
work does not capture the complexity of the preprocessing stage
(mostly cubic in the lattice dimension). In practice, this
assumption is justified in slowly varying channels where the
complexity of the preprocessing stage will be shared by many
transmission frames (e.g., a wired ISI channel or a wireless
channel with stationary terminals). If the number of these frames
is large enough, i.e., the channel is slow enough, the
preprocessing complexity can be ignored compared to the complexity
of the tree search stage which has to be independently performed
in every frame. Optimizing the complexity of the preprocessing
stage, however, is an important topic, especially for fast fading
channels.

\subsection{Taming the Channel: Left Preprocessing}\label{taming}

In the case of uncoded transmission ($\Gm = \Id$), QR
decomposition of the channel matrix $\Hm$ (assuming rank$(\Hm) =
m$) allows simple recursive detection of the information symbols
$\xv$. Indeed, ${\Qm}$ is the feedforward matrix of the
zero-forcing decision feedback equalizer (ZF-DFE) \cite{vblast}.
In general, sphere decoders can be seen as ZF-DFEs with some
reprocessing capability of their tentative decisions.

It is well-known that ZF-DFE is outperformed by the MMSE-DFE in
terms of signal-to-interference plus noise ratio (SINR) at the
decision point, under the assumption of correct decision feedback
\cite{forney-cioffi-mmsegdfe}. This observation motivates the
proposed approach for left preprocessing \cite{damen2}. This new
matrix can be obtained through the QR decomposition of the
augmented channel matrix
\begin{eqnarray} \label{augmentedH}
\tilde{\Hm} & \defines & \left[\begin{array}{c} \Hm  \\
\Id  \\
\end{array}\right]=\tilde{\Qm} \Rm_1
\end{eqnarray}
where $\tilde{\Qm} \in \RR^{(n+m) \times m}$ has orthonormal
columns and $\Rm_1$ is upper triangular. Let $\Qm_1$ be the upper
$n\times m$ part of $\tilde{\Qm}$. ${\Qm}_1$ and ${\Rm}_1$ are the MMSE-DFE forward and backward
filters, respectively. Thus, the transformed channel matrix and the received sequence are given by $\Rm_1$
and $\yv'=({\mb Q}_1^\transp{\rv}-\Rm_1 \vv)$,
respectively. The transformed CLPS
\begin{equation} \label{sys0a-mmse}
\min_{\xv \in \Uc} \Big| {\mb y}' - \Rm_1 \Gm \xv \Big|^2
\end{equation}
is  not equivalent to (\ref{sys0a}) since, in general, $\Qm_1$
does not have orthonormal columns. The additive noise $\wv =
\yv' - \Rm_1 \Gm\xv$ in (\ref{sys0a-mmse}) contains both a
Gaussian component, given by  $\Qm_1^\transp \zv$, and a
non-Gaussian (signal-dependent) component, given by
$(\Qm_1^\transp \Hm - \Rm_1) (\cv+\vv)$. Nevertheless, for lattice
codes such that cov$(\cv+\vv) = \Id$, it can be shown that
cov$(\wv) = \Id$ \cite{hesham1}. Hence, the additive noise
component $\wv$ in (\ref{sys0a-mmse}) is still white, although
non-Gaussian and data dependent. Therefore, the minimum distance rule
(\ref{sys0a-mmse}) is expected to be only slightly
suboptimal.\footnote{This argument can be made rigorous by
considering certain classes of lattices of increasing dimension,
Voronoi shaping and random uniformly distributed dithering common
to both the transmitter and the receiver, as shown in
\cite{hesham1,erez-zamir}.} On the other hand, the augmented
channel matrix $\widetilde{\Hm}$ in (\ref{augmentedH}) has always
rank equal to $m$ and it is well conditioned, since $\Rm_1^\transp
\Rm_1 = \Id + \Hm^\transp \Hm$. Therefore, in some sense we have
{\em tamed} the channel at the (small) price of the
non-Gaussianity of the noise. The better conditioning achieved by
the MMSE-DFE preprocessing is illustrated in Fig.\ref{lattices}
(b) and (c).

\subsection{Inducing Sparsity: Right Preprocessing} \label{lll_latt}

In order to obtain the tree structure, one needs to put $\mb{R}_1
\Gm$ in upper triangular form $\Rm$ via QR decomposition. The
sparser the matrix  ${\mb R}$, the smaller the complexity of the
tree search algorithm. For example, a  diagonal ${\mb R}$ means
that symbol-by-symbol detection is optimal, i.e., the tree search
reduces to exploring a single path in the tree. Loosely, if one
adopts a depth first search strategy, then a sparse ${\mb R}$ will
lead to a better {\em quality} of the first leaf node found by the
algorithm.\footnote{More details on the different search
strategies are reported in Section~\ref{search_algos}.}
Consequently, the algorithm finds the closest point in a shorter
time \cite{damen1}.

While we have no rigorous method for relating the ``sparsity'' of
$\Rm$ to the complexity of the tree search, inspired by decision
feedback equalization in ISI channel, we define the sparsity index
of the upper triangular matrix $\mb{R}$ as follows
\begin{eqnarray}\label{sparsity-index}
\sparse(\Rm) &\defines& \max_{i\in \{1, \hdots, m\}}
\frac{\sum_{j=i+1}^{m}r_{i,j}^2}{r_{i,i}^2}.
\end{eqnarray}
where $r_{i,j}$ denotes the $(i,j)$-th element of $\Rm$. One can
argue that the smaller $\sparse(\Rm)$ the sparser $\Rm$ (e.g.,
$\sparse(\Rm)=0$ for $\Rm$ diagonal). The goal of right
preprocessing is to find a change of basis of the lattice
$\{\Rm_1\Gm \xv : \xv \in \ZZ^m\}$, such that the new lattice
generator matrix, $\Sm$, satisfies $\Sm = \Qm \Rm$ with
$\sparse(\Rm)$ as small as possible. This amounts to finding a
unimodular matrix $\Tm$ (i.e., the entries of $\mb T$ and ${\mb
  T}^{-1}$ are integers) such that $\Rm_1 \Gm = \Qm \Rm \Tm$
with $\Qm$ unitary and $\sparse(\Rm)$ minimized over the group of
unimodular matrices. This optimization problem appears very
difficult to solve; however, there exist many heuristic approaches
to find unimodular matrices that give small values of
$\sparse(\Rm)$. Examples of such methods, considered here, are
lattice reduction, column permutation and a combination thereof.

Lattice reduction finds a reduced lattice basis, i.e., the columns
of the reduced generator matrix $\mb S$ have ``minimal'' norms and
are as orthogonal as possible.\footnote{For more details on the
different notions and methods of lattice reduction, the reader is
referred to \cite{cohen}.} The most widely used reduction
algorithm is due to Lenstra, Lenstra and Lov\'asz (LLL) \cite{LLL}
and has a polynomial complexity in the lattice dimension. An
enhanced version of the LLL algorithm, namely the deep insertion
modification, was later proposed by Schnorr and Euchner
\cite{schn-euch}. LLL with deep insertion gives a reduced basis
with significantly shorter vectors \cite{cohen}. In practice, the
complexity of the LLL with deep insertion is similar to the
original one even though it is an exponential time algorithm in
the worst case sense \cite{cohen}.

Another method for decreasing $\sparse(\Rm)$ consists of ordering
the columns of $\Rm_1 \Gm$, i.e., by right-multiplication by a
permutation matrix $\Sigmam$. In  the sequel, we shall use the
V-BLAST greedy ordering strategy proposed in
\cite{vblast,fast-vblast}. This algorithm finds a permutation
matrix $\Sigmam$ such that $\Rm_1\Gm = \Qm \Rm \Sigmam$ maximizes
$\min_{i} r_{i,i}^2$. Since $\Rm^\transp \Rm = \Sigmam^-{\transp}
\Gm^\transp \Rm_1^\transp \Rm_1 \Gm \Sigmam^{-1}$, i.e., the set
$\{\sum_{j} r_{i,j}^2 \; : \; i=1,\ldots,m\}$ depends only on
$\Rm_1\Gm$ and not on $\Sigmam$, by maximizing the minimum
$r^2_{i,i}$ this algorithm miminizes $\sparse(\Rm)$ over the group
of permutation matrices (a subgroup of the unimodular matrices).

Lattice reduction and column permutation can be combined. This
yields an unimodular matrix $\Tm = \Sigmam \Tm_1$, where $\Tm_1$
is obtained by lattice-reducing $\Rm_1\Gm$ and $\Sigmam$ by
applying the V-BLAST greedy algorithm on the resulting reduced
matrix $\Rm_1\Gm \Tm_1^{-1}$.

As observed before, the unimodular right multiplication does not
change the lattice but may significantly complicate the boundary
control. In fact, we have
\begin{eqnarray} \label{sys0a-mmse-red}
\min_{\xv \in \Uc} \Big| {\mb y}' - \Rm_1 \Gm \xv \Big|^2 & = &
\min_{\xv \in \Uc} \Big| {\mb y}' - \Qm \Rm \Tm \xv \Big|^2 \nonumber \\
& = & \min_{\xv \in \Tm \Uc} \Big| \Qm^\transp {\mb y}' - \Rm \xv
\Big|^2
\end{eqnarray}
The new constraint set $\Tm \Uc$ might be even more complicated to
enforce than the original information set $\Uc$ (see
Fig.~\ref{lattices}(d)). However, it is clear that although
modifying the boundary control may result in a significant
complexity increase for ML decoding, lattice decoding is not
affected at all, since $\Tm \ZZ^m = \ZZ^m$.

\subsection{Forming the Tree} \label{qr}

\begin{figure}[!h]   % fig 1
\begin{center}
\epsfig{file=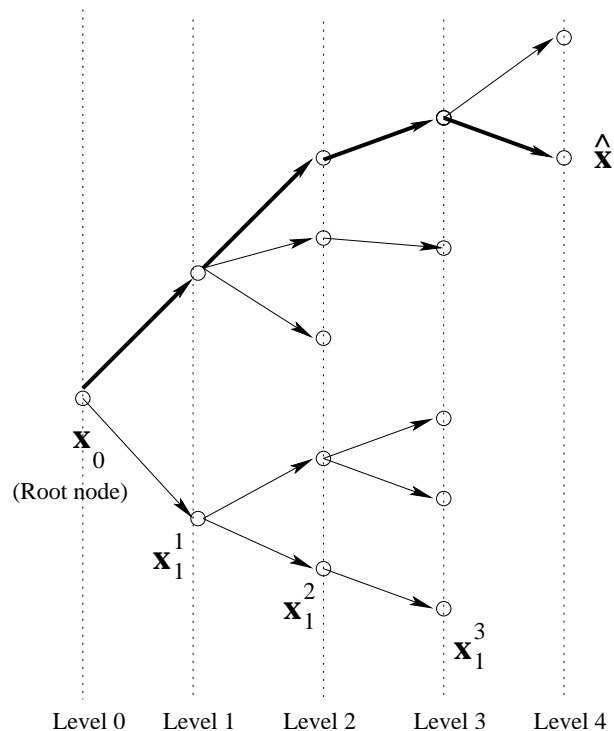,width=8.0cm}
\end{center}
\caption{Tree representation of the
               paths searched by sequential decoding algorithms
               in the case $m=4$.}
\label{tree-rep}
\end{figure}

\begin{figure}
\begin{center}
\begin{tabular}{cc}
\includegraphics[width=2in,height=2in]{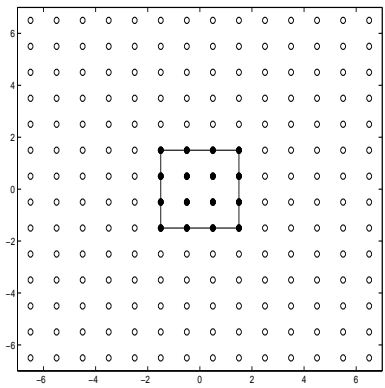} &
\includegraphics[width=2in,height=2in]{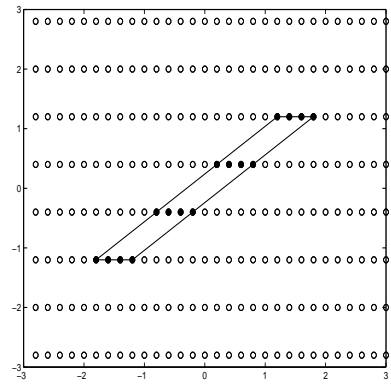}  \\
{\footnotesize (a) The translated $\ZZ^2$ lattice and QAM constellation} &
{\footnotesize (b) The received lattice
after channel distortion (constellation)}\\
\includegraphics[width=2in,height=2in]{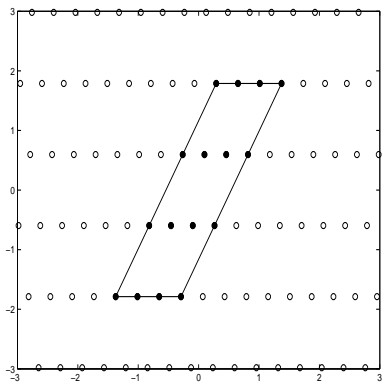} &
\includegraphics[width=2in,height=2in]{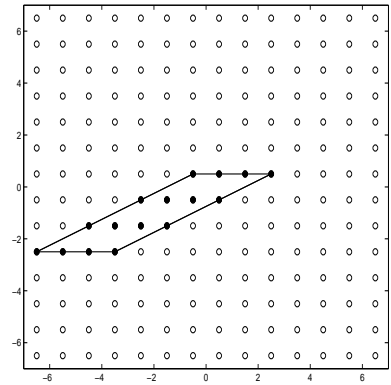} \\
{\footnotesize (c) After MMSE-DFE left preprocessing} & {\footnotesize (d) Boundary control after right preprocessing}\\
\end{tabular}
\caption{The effect of left preprocessing on the lattice and the right
preprocessing on the information set}
\label{lattices}
\end{center}
\end{figure}

The final step in preprocessing is to expose the tree structure of
the  problem. In this step, QR decomposition is applied on the
transformed combined channel and lattice matrix $\Qm_1^\transp
\Hm\Gm \Tm^{-1}$, after left and right preprocessing. The upper
triangular nature of ${\mb R}$ means that a tree search can now be
used to solve the CLPS problem. Fig. \ref{tree-rep} illustrates an
example of such a tree.

%%%%%%%%%%%%%%%%%%%%%%%%%%
Here, we wish to stress that our approach for exposing the tree is
fundamentally different from the one traditionally used for codes
over finite alphabets (e.g., linear block codes, convolutional
codes, trellis coset codes in AWGN channels). Here, we operate over the field of
real numbers and consider the lattice corresponding to the joint
effect of encoding and channel distortion. In the conventional
approach, the tree is generated from the trellis structure of the
code alone, and hence, does not allow for a natural tree search
that handles jointly detection (the linear channel) and decoding.
In fact, joint detection and decoding is achieved at the expenses
of an increase of the overall system memory (joint trellis), or by
neglecting some paths in the search (e.g., by per-survivor reduced
state processing). Since operating on the full joint trellis is
usually too complex, both the proposed and the conventional
per-survivor (reduced state) approach are suboptimal, and the
matter is to see which one achieves the best
performance/complexity tradeoff.
%%%%%%%%%%%%%%%%%%%%%%%%%%

For the sake of convenience, in the following we shall denote
again by $\yv$ the channel output after all transformations, i.e.,
the tree search is applied to the CLPS problem $\min_{\xv \in
\ZZ^m} |\yv - \Rm \xv|^2$ with $\Rm$ in upper triangular form. The
components of vectors and matrices are numbered in reverse order,
so that the preprocessed received signal can finally be written as
\begin{equation} \label{notation}
\left(\begin{array}{c} y_m \\ \vdots \\ y_{1} \end{array}
\right)=\left(\begin{array}{cccc}
r_{m,m} & \hdots & \hdots & r_{m,1} \\
0 & r_{m-1,m-1} & \hdots & r_{m-1,1} \\
\vdots  & \ddots & \ddots & \vdots \\
0& \hdots & 0 & r_{1,1}\end{array}\right)\left(\begin{array}{c} x_m \\
\vdots \\ x_{1} \end{array}\right) + \left(\begin{array}{c} w_m \\
\vdots \\ w_{1} \end{array}\right).
\end{equation}
Notice that after preprocessing the problem is always squared, of
dimension $m$, even though the original problem has arbitrary $m$
and $n$. Throughout the paper, we consider a tree rooted at a
fixed dummy node $x_0$. The node at level $k$ is denoted by the
label ${\mathbf x}_1^k= (x_1,x_2,...,x_k)$. Moreover, every node
$\xv_1^k$ is associated with the the squared distance
\begin{equation} \label{wi}
w_k({\xv}_1^k) = \left|y_k - \sum_{j=1}^{k} r_{k,j}x_j \right|^2.
\end{equation}

The difference between the transmitted codeword $\hat{\mb x}$ and
any valid codeword ${\mb x}$ is denoted by $\tilde{\mb x}$, i.e.,
$\tilde{\mb x} = \hat{\mb x} - {\mb x}$.

We hasten to stress that the preprocessing steps highlighted in
Sections \ref{taming}-\ref{qr} are for a general setting. In some
special cases, some steps can be eliminated or alternative options
can be used. Some of these cases are listed hereafter.
\begin{enumerate}
\item {\em Upper Triangular Code Generator Matrix}:\\
In this case, after taming the channel, $\mb{H} \longrightarrow
\mb{R}_1$, the new combined matrix $\mb{R}_1 \mb{G}$ is also upper
triangular and can be directly used to form the tree without any
further preprocessing (if one decides against right
preprocessing).
\item {\em Uncoded V-BLAST}: \\
For uncoded V-BLAST systems (i.e., $\mb{G} = \mb{I}$), applying
the MMSE-DFE greedy ordering of
\cite{hassibi-squareroot,fast-vblast} may achieve better
complexity of the tree search stage than applying MMSE-DFE left
preprocessing, lattice reduction, and greedy ordering of the final
QR decomposition. This is especially true for large dimensions,
where lattice reduction is less effective \cite{damen1}.

\item {\em The Hermite Normal Form Transformation}:\\
Ultimately, any hardware implementation of the decoder requires
finite arithmetics. In this case, all quantities are scaled and
quantized such that they take on integer values. While all the
preprocessing steps in Sections \ref{taming}-\ref{qr} can be
easily adapted to finite arithmetics, there exist other efficient
transformations for integral matrices that may yield smaller
complexity over the ones mentioned above. For example, one can
apply the Hermite normal form (HNF)  \cite{cohen} directly on the
scaled (and quantized) matrix $\tau \Hm \Gm$, such $\tau \Hm \Gm =
\Rm \Tm$, with $\Tm$ unimodular and $\Rm$ upper triangular with
the property that each diagonal element dominates the rest of the
entries on the same row (i.e., $r_{i,i} > r_{i,j} \geq 0, \,
i=1,\ldots,m, \, j=i+1,\ldots, m$). Interestingly, the HNF
transformation improves the sparsity index and reduces the
preprocessing to a single step.

\end{enumerate}

%-----------------------------------------------------------------------------------
%-----------------------------------------------------------------------------------

\section{The Tree Search Stage}\label{search_algos}
After proper preprocessing, the second stage of the CLPS
corresponds to an instance of searching for the best path
in a tree. In this setting, the tree has a maximum depth $m$, and
the goal is to find the node(s) at level $m$ that has the least
squared distance, where the squared distance for any node ${\mb
x}_1^m$ at level $m$ (called {\it leaf} node) is given by
\begin{eqnarray}
d^2({\xv},{\xv}_1^m)&=&\sum_{i=1}^m w_i({\xv}_1^i)
\end{eqnarray}
Visiting all leaf nodes to find the one with the least metric, is
either prohibitively complex (exponential in $m$), or not
possible, as with lattice decoding. The complexity of tree search
can be reduced by the branch and bound (BB) algorithm which
determines if an intermediate node ${\mb x}_1^k$, on extending,
has any chance of yielding the desired leaf node. This decision is
taken by comparing the {\it cost function} assigned to the node by
the search algorithm, against a {\it bounding function}. In the
following section, we propose a generic tree search stage,
inspired by the BB algorithm, that encompasses many known
algorithms for CLPS as its special cases. We further use this
algorithm to classify various tree search algorithms and elucidate
some of their structural properties.

\subsection{Generic Branch and Bound Search Algorithm}\label{bb-sec}
Before describing the proposed algorithm, we first need to
introduce some more notation.

\begin{itemize}
\item ACTIVE is an ordered list of nodes.

\item $f({\mb x}_1^k)\in {\mathbb R}$ is the {\it cost function}
of any node ${\mb x}_1^k$ in the tree, and ${\mb t}\in {\mathbb
R}^{m\times 1}$ is the bounding function.

\item Any node ${\mb x}_1^k$ in the search space of the search
algorithm is a {\it valid} node, if $f({\mb x}_1^k)<t_k$.

\item A node is {\it generated} by the search algorithm, if the
node occupies any position in ACTIVE at some instant during the
search.

\item ``{\it sort}'' is a rule for ordering the nodes in the list
ACTIVE.

\item ``{\it gen}'' is a rule defining the order for generating
the child nodes of the node being extended.

%\item ``{\it prune}'' is a rule for pruning the ACTIVE list.

\item $g_1$ and $g_2$ are rules for tightening the bounding
function.

\item At any instant, the leaf node with the least distance
generated by the search algorithm so far in the search process is
stored in $\hat{\mb x}$.

\item We define the search {\it complexity} of a tree search
algorithm as the number of nodes generated by the algorithm.

\item Two search algorithms are said to be equivalent if they
generate the same set of nodes.

\item A BB algorithm whose solution is guaranteed to be (one of)
leaf node(s) with least distance is called an {\it optimal} BB
algorithm. If the solution is not guaranteed to have the least
distance to ${\mb y}$ among all leaf nodes, then the BB algorithm
is a {\it heuristic} BB algorithm.

\end{itemize}

We are now ready to present our generic search algorithm.\\

{\bf GBB($f$, ${\mb t}$, {\it sort}, {\it gen}, $g_1$, $g_2$)}:

\begin{enumerate}

\item Create the empty list ACTIVE, and place the root node in
ACTIVE. Set $n_c\leftarrow 1$.

\item Let ${\mb x}_1^k$ be the top node of ACTIVE.

{\it If} ${\mb x}_1^k$ is a leaf node ($k=m$), then

\hspace{0.1in} ${\mb t}\leftarrow g_1({\mb t}, f({\mb x}_1^m))$
and $\hat{\mb x}\leftarrow \arg \min( \sum_{i=1}^m w_i({\mb
x}_1^i),\sum_{i=1}^m w_i(\hat{\mb x}_1^i))$.

\hspace{0.1in} Remove ${\mb x}_1^m$ from ACTIVE.

\hspace{0.1in} {\it Go to} step~4.

{\it If} ${\mb x}_1^k$ is not a valid node, then remove ${\mb
x}_1^k$ from ACTIVE. {\it Go to} step~4.

{\it If} all valid child nodes of ${\mb x}_1^k$ have already been
generated, then remove ${\mb x}_1^k$ from ACTIVE. {\it Go to}
step~4.

Generate a valid child node ${\mb x}_1^{k+1}$ of ${\mb x}_1^k$, not generated
before, according to the order {\it gen}, and place it in ACTIVE.
Set $n_c\leftarrow n_c+1$. Set ${\mb t}\leftarrow g_2({\mb
t},n_c,\mbox{ACTIVE})$. Update $f(\xv_1^k),f(\xv_1^{k+1})$.

\item Sort the nodes in ACTIVE according to {\it sort}.

\item {\it If} ACTIVE is empty, then exit. {\it Else, Go to} step~2.

\end{enumerate}

In {\bf GBB}, $g_1$ allows one to tighten the bounding function
when a leaf node reaches the top of ACTIVE, whereas $g_2$ allows
for restricting the search space in heuristic BB algorithms. For
example, setting
\[g_2({\bf t},n_{c,t},\mbox{ACTIVE})=\left[-\infty,-\infty,...,-\infty\right]^\transp,\]
will force the search algorithm to terminate when the number of
nodes generated increases beyond a tolerable limit on the
complexity given by $n_{c,t}$. Whenever a leaf node reaches the
top of ACTIVE, $\hat{\mb x}$ is updated if appropriate. Now, we
use {\bf GBB} to classify various tree search algorithms in three
broad categories. This classification highlights the structural
properties and advantages/disadvantages of the different search
algorithms.

\subsubsection{Breadth First Search} \label{brfs}

{\bf GBB} becomes a Breadth First Search (BrFS) if $g_1({\mb
t},f({\mb x}_1^m))={\mb t}$, and the cost function $f$ of any node, once
determined, is never updated.
Ultimately, all nodes ${\mb x}_1^k$ whose cost function along the
path ${\mb x}_1^k$ does not rise above the bounding function, are
generated before the algorithm terminates, unless
the $g_2$ function removes their parent nodes from ACTIVE. Now, we
can establish the equivalence between various sphere/sequential
decoders and BrFS.

The first algorithm is the Pohst enumeration strategy reported in
\cite{pohst}. In this strategy, the bounding function ${\mb t}$
consists of equal components $C_0$, where $C_0$ is a constant
chosen before the start of search\footnote{For the sake of
simplicity, we assumed in the above classification that the
bounding function is chosen such that at least one leaf node is
found before the search terminates. If, however, no leaf node is
found before the search terminates, the bounding function is
relaxed and the search is started afresh.}, and the cost function
of a node ${\xv}_1^k$ is $f({\xv}_1^k)=\sum_{i=1}^k
w_i({\xv}_1^i)$. Therefore, all nodes ${\xv}_1^k$ in the search
space that satisfy
\begin{equation} \label{pohst_enum_1}
\sum_{i=1}^k w_i({\mb x}_1^i)\leq C_0
\end{equation}
are generated before termination. Generating the child nodes in
this strategy is simplified by the following observation. For any
parent node ${\xv}_1^k$, the condition
$\displaystyle{\sum_{i=1}^{k+1} w_i({\xv}_1^i)\leq C_0}$ for the
set of generated child nodes implies that the $(k+1)-$th component
of the generated child nodes lies in some interval $[a_0,a_1]$.
The second example is the statistical pruning (SP) decoder which
is equivalent to a heuristic BrFS decoder. Two variations of SP
are proposed in \cite{radhika1}, the increasing radii (IR) and
elliptical pruning (EP) algorithms. The IR algorithm is a BrFS
with the bounding function ${\mb t}=\{t_1,...,t_m\}$, where $t_k,
1\leq k\leq m$ are constants chosen before the start of search.
The cost function for any node in IR is the same as in Pohst
enumeration. The EP algorithm is given by the bounding function
${\mb t}=\{1,...,1\}$, and the cost function for the node ${\mb
x}_1^k$ given by $\displaystyle{f({\mb x}_1^k)=\sum_{i=1}^{k}
\frac{w_i({\mb x}_1^i)}{e_k}}$, where $e_k, 1\leq k\leq m$ are
constants. More generally, when $g_2(.)={\mb t}$, i.e., $g_2$ is
not used, the resulting BrFS algorithm is equivalent to the
Wozencraft sequential decoder \cite{wozencraft} where, depending
on the cost function, the decoder can be heuristic or optimal.

The $M$ algorithm \cite{anderson} and $T$-algorithm \cite{kokkonen} are also examples of heuristic BrFS.
Here, however, $g_2$ serves an important role in restricting the
search space. In both algorithms, {\it sort} is defined as
follows. Any node in ACTIVE at level $k$ is placed above any node
at level $k+1$, and nodes in the same level are sorted in
ascending order of their cost functions. In the $M$-algorithm,
after the first node at level $k+1$ is generated (indicating that
all valid nodes at level $k$ have already been generated), $g_2$
sets $t_k$ to the cost function of the $M$-th node at level $k$
(where $M$ is an initial parameter of the $M$-algorithm). In the
$T$-algorithm, $g_2$ sets $t_k$ to $(f(\bar{\mb x}_1^k)+T)$, where
$\bar{\mb x}_1^k$ is the top node at level $k$ in ACTIVE, and $T$
is a parameter of the $T$-algorithm. After $t_k$ is tightened in
this manner, all nodes in ACTIVE at level $k$, that satisfy
$f({\mb x}_1^k)>t_k$ are rendered invalid, and are subsequently
removed from ACTIVE.

%{\bf Add discussion of the M-algorithm and the Wozencraft
%decoder.\\}
In general, BrFS algorithms are naturally suited for applications
that require soft-outputs, as opposed to a hard decision on the
transmitted frame. The reason is that such algorithms output an
ordered\footnote{The list is ordered based on the cost function of
the different candidates} list of candidate codewords. One can
then compute the soft-outputs from this list using standard
techniques (e.g., \cite{hoch-ten},\cite{hagenauer}). Here, we note that in the proposed
joint detection and decoding framework, soft outputs are generally not needed.
Another advantage of BrFS is
that the complexity of certain decoders inspired by this strategy
is robust against variations in the SNR and channel conditions.
For example, the $M$-algorithm has a constant complexity
independent of the channel conditions. This property is appealing
for some applications, especially those with hard limits on the
maximum, rather than average, complexity. On the other hand,
decoders inspired by the BrFS strategy usually offer poor results
in terms of the average complexity, especially at high SNR. One
would expect a reduced average complexity if the bounding function
is varied during the search to exploit the additional information
gained as we go on. This observation motivates the following
category of tree search algorithms.

%-----------------------------------------------------------------------------------
\subsubsection{Depth First Search} \label{dfs}

{\bf GBB} becomes a depth first search (DFS) when the following
conditions are satisfied. The sorting rule {\it sort} orders the
nodes in ACTIVE in reverse order of generation, i.e., the last
generated node occupies the top of ACTIVE, and
\[g_1({\mb t},f({\mb x}_1^m))=\left[\min(t_1,f({\mb
    x}_1^m)),...,\min(t_m,f({\mb x}_1^m))\right]^\transp.\] As in BrFS, the
cost function of any node, once generated, remains constant. Even among
algorithms within the class of DFS algorithms, other
parameters, like {\it gen} and $g_2$, can significantly alter the
search behavior. To illustrate this point, we contrast in the
following several sphere decoders which are equivalent to DFS
strategies.

The first example of such decoders is the modified Viterbo-Boutros
(VB) decoder reported in \cite{damen1}. In this decoder, $g_2({\mb
t},n_c)={\mb t}$, and the cost function for any node ${\mb x}_1^k$
is $f({\mb x}_1^k)=\sum_{i=1}^k w_i({\mb x}_1^i)$. For any node
${\mb x}_1^k$ and its corresponding interval $[a_0,a_1]$ for valid
child nodes, the function {\it gen} generates the child node with
$a_0$ as its $(k+1)^{th}$ component first. Our second example is
the Schnorr Euchner (SE) search strategy first reported in
\cite{agrell}. This decoder shares the same cost functions and
$g_2$ with the modified VB decoder, but differs from it in the
order of generating the child nodes. For any node ${\mb x}_1^k$
and its corresponding interval $[a_0,a_1]$ for the valid child
nodes, let $a_{m}\defines\lfloor\frac{a_0+a_1}{2}\rceil$
and $\delta\defines\sign(w_{k+1}({\mb x}_1^{k+1}))$. Then, the
function {\it gen} in the SE decoder generates nodes according to
the order
$\{a_{m},a_{m}+\delta,a_{m}-\delta,a_{m}+2\delta,
\hdots\}$.

Due to the adaptive tightening of the bounding function, DFS
algorithms have a lower average complexity than the corresponding
BrFS algorithms with the same cost functions, especially at high
SNR. Another advantage of the DFS approach is that it allows for
greater flexibility in the performance-complexity tradeoff through
carefully constructed termination strategy. For example, if we
terminate the search after finding the first leaf node, i.e.,
$n_c=m$,  then we have the MMSE-Babai point decoder \cite{damen2}. This
decoder corresponds to the MMSE-DFE solution aided with the right
preprocessing stage. It was shown in \cite{damen2} that the
performance of this decoder is within a fraction of a dB from the
ML decoder in systems with small dimensions. The fundamental
weakness of DFS algorithms is that the sorting rule is {\em
static} and does not exploit the information gained thus far to
speed up the search process.

\subsubsection{Best First Search}

{\bf GBB} becomes a best first search (BeFS) when the following
conditions are satisfied. The nodes in ACTIVE are sorted in
ascending order of their cost functions, and
\[g_1({\mb t},f({\mb x}_1^m))=\left[\min(t_1,f({\mb x}_1^m)), \hdots,\min(t_m,f({\mb x}_1^m))\right]^\transp.\]
Note that in BeFS, the search can be terminated once a leaf node
reaches the top of the list, since this means that all
intermediate nodes have cost functions higher than that of this
leaf node. Thus, the bounding function is tightened just once in
this case.
The stack algorithm is an example of BeFS
decoder obtained by setting $g_2({\mb t},n_c)={\mb t}$, and the
cost function of any node in ACTIVE at any instant defined as follows: If
$\xv_1^k$ is a leaf node, then $f(\xv_1^k)=-\infty$. Otherwise, we let $\xv_{1,g}^{k+1}$ be the best child node of
$\xv_1^k$ not generated yet, and define $f(\xv_1^k)=\sum_{i=1}^{k+1}
w_i(\xv_{1,g}^{k+1})-b(k+1)$, where we refer to $b\in {\mathbb R}^+$ as the {\it bias}.
Because of
the efficiency of the sorting rule, BeFS algorithms are generally
more efficient than the corresponding BrFS and DFS algorithms.
This fact is formalized in the following theorems.
Theorem~1 establishes the efficiency of the stack
decoder with $b=0$ among all known sphere decoders.

\begin{theorem} \cite{Xu} \label{thm_bfs1}
The stack algorithm with $b=0$ generates the least number of nodes among
all optimal tree search algorithms.
\end{theorem}

The following result compares the heuristic stack algorithm, {\it i.e.,}
$b>0$, with a special case of the IR algorithm \cite{radhika1}, where the
bounding function takes the form $t_k=bk+\delta$.

\begin{theorem} \label{thm_bfs2}
The IR algorithm with cost function $\{\tv:t_k=bk+\delta\}$, generates at least as many nodes as those
generated by the stack algorithm when the same bias $b$ is used.
\end{theorem}

{\it Proof: } Appendix~\ref{bfs_vs_dfs}

At this point, it is worth noting that in our definition of search complexity, we count only the number of
generated nodes, {\it i.e.,} nodes that occupy some position in ACTIVE at
some instant. In general, this is a reasonable abstraction of the
actual computational complexity involved. However, in the stack algorithm,
for each node generated, the cost functions of two nodes are
updated instead of one; one for the generated node, and one for the parent node. Thus,
the comparisons in Theorems~\ref{thm_bfs1} and \ref{thm_bfs2} are
not completely fair.

Finally, we report the following two advantages offered by the the
stack algorithm. First, it offers a natural solution for the
problem of choosing the initial radius (or radii), which is
commonly encountered in the design of sphere decoders (e.g.,
\cite{damen1}). By setting all the components of $\mb t$ to
$\infty$, it is easy to see that we are guaranteed to find the
closest lattice point while generating the minimum number of nodes
(among all search algorithms that guarantee finding the closest
point). Second it allows for a systematic approach for trading-off
performance for complexity. To illustrate this point, if we set $b=0$, we
obtain the closest point lattice decoder (i.e., best performance
but highest complexity). On the other extreme, when
$b\rightarrow\infty$, the stack decoder reduces to the MMSE-Babai point
decoder discussed in the DFS section (the number of nodes visited
is always equal to $m$). In general, for systems with small $m$,
one can obtain near-optimal performance with a relatively large
values of $b$. As the number of dimensions increases, more
complexity must be expended (i.e., smaller values of $b$) to
approach the optimal performance.

\subsection{Iterative Best First Search}
In Section~\ref{bb-sec}, our focus was primarily devoted to
complexity, defined as the number of nodes visited by the tree
search algorithm. Another important aspect is the memory
requirement entailed by the search. Straightforward implementation
of the {\bf GBB} algorithm requires maintaining the list ACTIVE,
which can have a prohibitively long length in certain application.
This motivates the investigation of {\em modified} implementations
of these search strategies that are more efficient in terms of
storage requirements. The BrFS and DFS sphere decoders discussed
in Sections~\ref{brfs}~and~\ref{dfs} lend themselves naturally to
storage efficient implementations. Such implementations have been
reported in \cite{pohst,damen1,damen2,agrell,radhika1}.

In order to exploit the complexity reduction offered by BeFS
strategy in practice, it is therefore important to seek modified
memory-efficient implementations of such algorithms. This can be
realized by storing only one node at a time, and allowing nodes to
be visited more than once. The search in this case progresses in
contours of increasing bounding functions, thus allowing more and
more nodes to be generated at each step, finally terminating once
a leaf node is obtained. The Fano decoder \cite{fano} is the {\em
iterative} BeFS variation of the stack algorithm. Although the
stack algorithm and the Fano decoder, with the same cost
functions, generate essentially the same set of nodes
\cite{geist}, the Fano decoder visits some nodes more than once.
However, the Fano decoder requires essentially no memory, unlike
the stack algorithm. Appendix~\ref{app_fano_algo} provides an
algorithmic description of the Fano decoder and a brief
description of the relevant parameters. Overall, the proposed
decoder consists of left preprocessing (MMSE-DFE) and right
preprocessing (combined lattice reduction and greedy ordering),
followed by the Fano (or stack) search stage for lattice, not ML,
decoding.

%%%%%%%%%%%%%%%%%%%%%%%%%%%%%%%%%%%%%%%%%%%%%%%%%%%%%%%%%%%%%%%%%%%%%%%%%%
\section{Analytical and Numerical Results} \label{analysis}
To illustrate the efficiency and generality of the proposed
framework, we utilize it in three distinct scenarios. First, we
consider uncoded transmission over MIMO channels (i.e., V-BLAST).
Here, we present analytical, as well as simulation, results that
demonstrate the excellent performance-complexity tradeoff achieved
by the proposed Stack and Fano decoders. Then, we proceed to coded
MIMO systems and apply tree search decoding to two different
classes of space-time codes. Finally, we conclude with trellis
coded transmission over ISI channels.
\subsection{The V-BLAST Configuration}
Unfortunately, analytical characterization of the performance-
complexity tradeoff for sequential/sphere decoders with arbitrary
${\mathbf H}{\mathbf G}$ and $\Uc$ still appears intractable. To
avoid this problem, we restrict ourselves in this section to
uncoded transmission over flat Rayleigh MIMO channels. In our
analysis, we further assume that ZF-DFE pre-processing is used.
The complexity reductions offered by the proposed preprocessing
stage are demonstrated by numerical results.

\begin{theorem} \label{thm_diversity}
The Stack algorithm and the Fano decoder with any finite bias $b$,
achieve the same diversity as the ML decoder when applied
to a V-BLAST configuration.
\end{theorem}

\begin{proof}
Appendix~\ref{proof_of_diversity}.
\end{proof}

The result shows that the Fano decoder, unlike other heuristic
algorithms like nulling-and-canceling, does not lead to a lower
diversity than the ML decoder.

\begin{theorem} \label{thm_stack_comp_2}
In a V-BLAST system with $Q^2$-QAM, the average complexity per
dimension of the stack algorithm for a sufficiently large bias $b$
is linear in $m$ when the SNR $\rho$ grows linearly with $m$ and
$r=n-m\geq 0$.
\end{theorem}

\begin{proof}
Appendix~\ref{stack_complexity}.
\end{proof}

\begin{figure}   % fig 1
\centerline{\subfigure{\epsfig{file=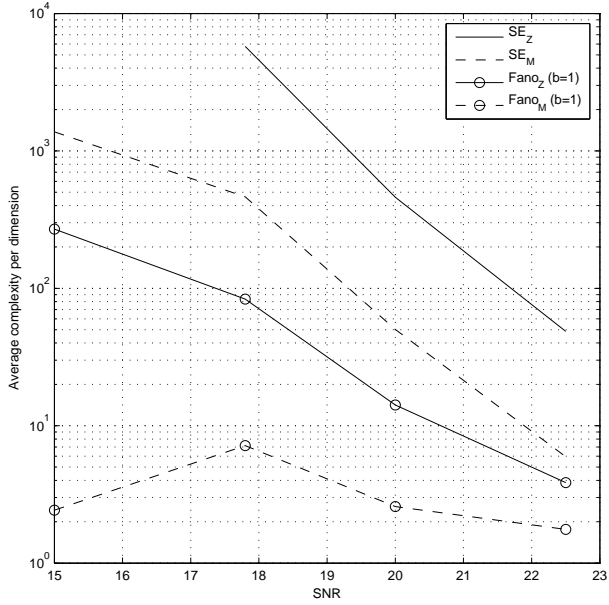,width=8.0cm}}
\hfill \subfigure{\epsfig{file=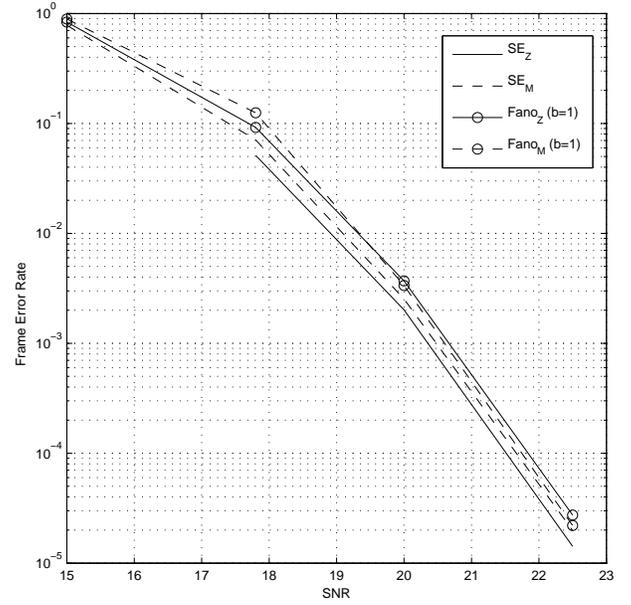,width=8.0cm}}}
\caption{Complexity and Performance of SE enumeration and Fano
decoder for a $20\times 20$ $16-$QAM V-BLAST system}
\label{figure1}
\end{figure}

\begin{figure}   % fig 2
\centerline{\subfigure{\epsfig{file=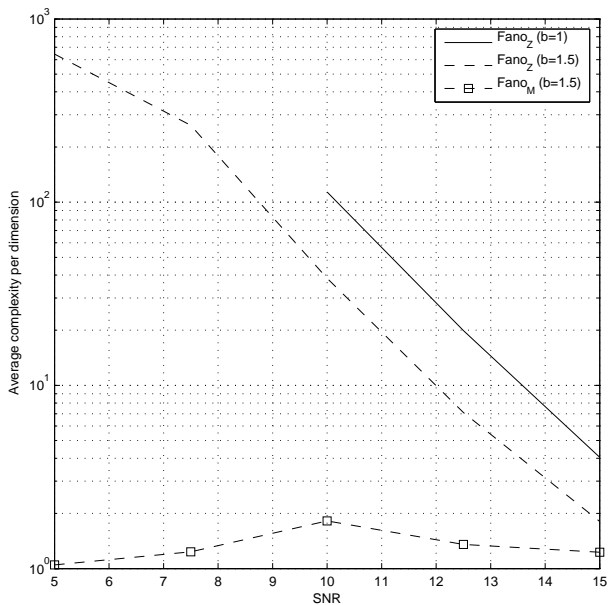,width=8.0cm}}
\hfill \subfigure{\epsfig{file=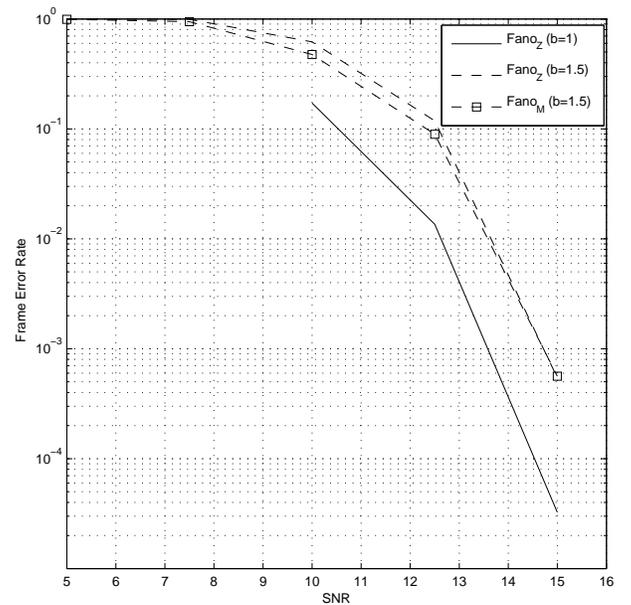,width=8.0cm}}}
\caption{Complexity and Performance of Fano decoder with ZF-DFE
and MMSE-DFE based preprocessing for a $30\times 30$ $4-$QAM
V-BLAST system} \label{figure2}
\end{figure}

\begin{figure}   % fig 3
\centerline{\subfigure{\epsfig{file=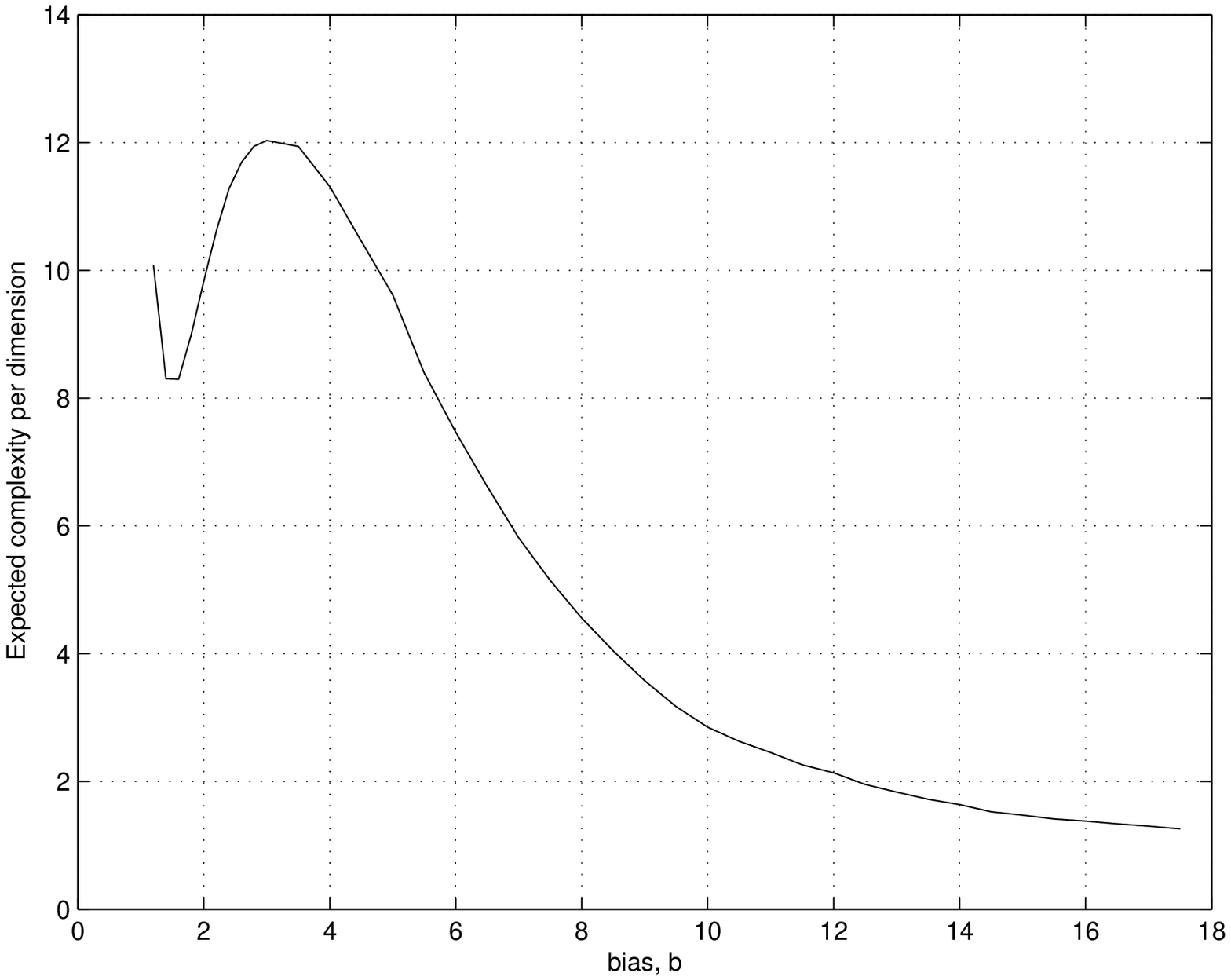,width=8.0cm}}
\hfill \subfigure{\epsfig{file=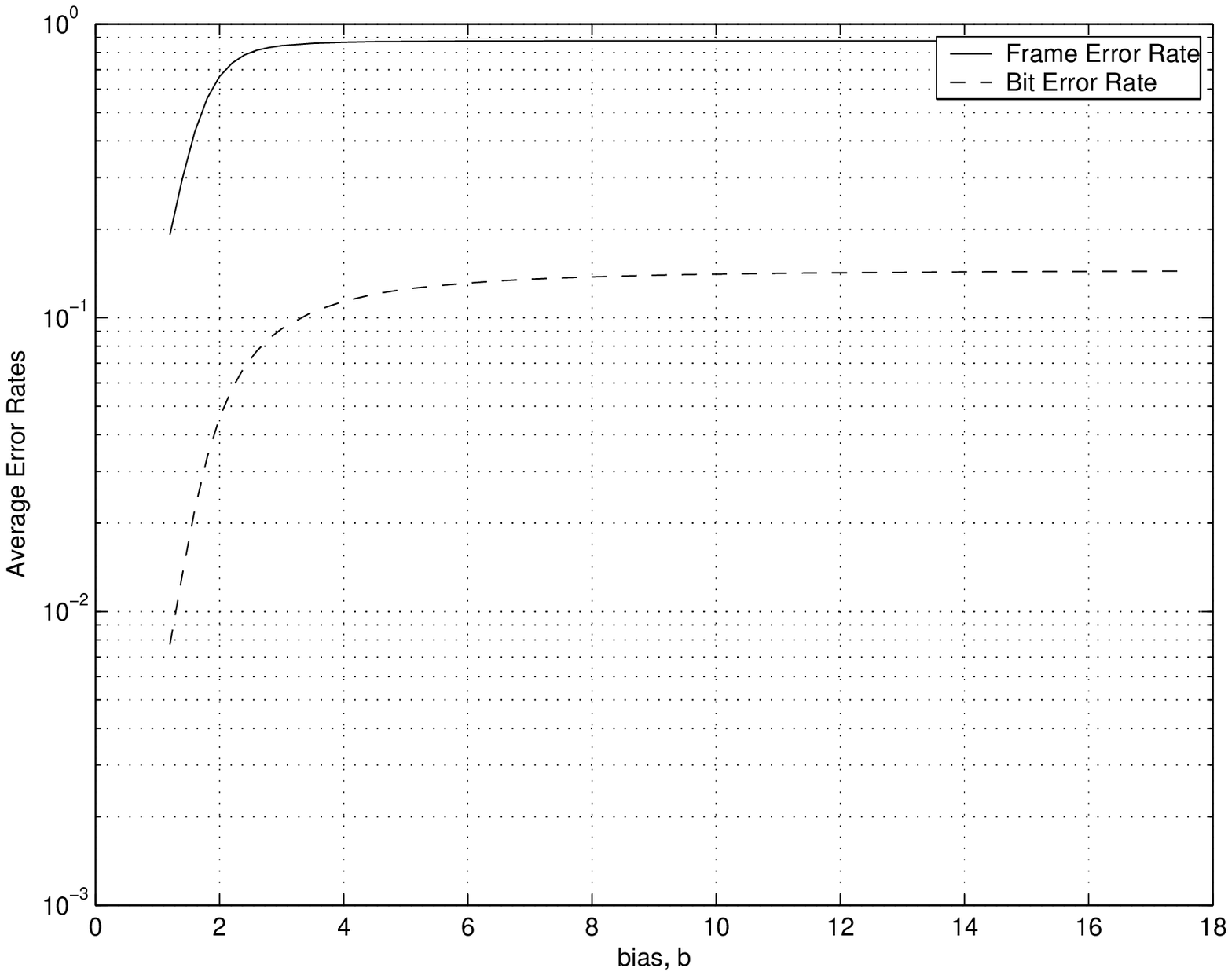,width=8.0cm}}}
\caption{Complexity and Performance of Fano decoder with different 
bias, for a $20\times 20$ $4-$QAM V-BLAST system with ZF-DFE
preprocessing} \label{figure3}
\end{figure}

\begin{figure}   % fig 4
\begin{center}
\epsfig{file=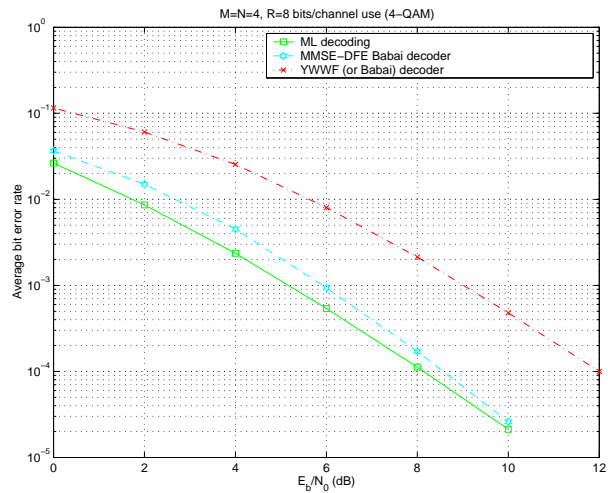,width=8.0cm}
\end{center}
\caption{Performance of MMSE-DFE preprocessing with DFE for a
$4\times 4$, $4-$QAM V-BLAST system} \label{figure4}
\end{figure}

Thus, one can achieve linear complexity with the stack algorithm
by allowing the SNR to increase linearly with the lattice dimension.
To validate our theoretical claims, we further report numerical
results in selected scenarios. In our simulations, we assume that
the channel matrix is square and choose the SE enumeration as the
reference sphere decoder for comparison purposes. In all the figures, the
subscript $Z$ refers to ZF-DFE left preprocessing and the subscript $M$
denotes MMSE-DFE left preprocessing followed by LLL reduction and V-BLAST greedy
ordering for right preprocessing. In Fig.~\ref{figure1}, the
average complexity per lattice dimension and frame error rate of
Fano decoder with $b=1$ and the SE sphere decoder are shown for
different values of SNR in a $20\times 20$ $16-$QAM V-BLAST
system. Thus, for $m=40$, the Fano decoder can offer a reduction
in complexity up-to a factor of 100. Moreover, the performance of
the the Fano decoder is seen to be only a fraction of a dB away from that of
the SE decoder, which achieves ML performance. We also see that
the frame error rate curves for both the Fano decoder and the SE
(ML) decoder have the same slope in the high SNR region, as
expected from our analysis. Fig.~\ref{figure2} compares the
complexity and performance of the Fano decoder with ZF-DFE and
MMSE-DFE based preprocessing, respectively, in a $30\times 30$
$4-$QAM V-BLAST system (i.e., $m=60$). From the figures, we see
that the MMSE-DFE based preprocessing plays a crucial role in
lowering the search complexity of the Fano decoder, despite the
apparent increase in search space due to lattice decoding.
Fig.~\ref{figure3} reports the dependence of the complexity of the
Fano decoder on the value of $b$. The complexity attains a local
minimum for some $b^*>1$, and for large values of $b$, the
complexity of the Fano decoder decreases as $b$ is increased. The
error rate, however, increases monotonically with $b$ and
approaches that of the MMSE-DFE Babai decoder as $b\rightarrow \infty$. For
small dimensions, the performance of the MMSE-DFE based Babai
decoder is remarkable. This DFS decoder terminates after finding
the first leaf node. Fig.~\ref{figure4} compares the performance
of this decoder with the ML performance for a $4\times 4$, $4-$QAM
V-BLAST system. We also report the performance of the Yao-Wornell
and Windpassinger-Fischer (YWWF) decoder which has the same
complexity as the MMSE-DFE Babai decoder \cite{wornell,windpass}.
It is shown that the performance of the proposed decoder is within
a fraction of a dB from that of ML decoder, whereas the algorithm
in \cite{wornell,windpass} exhibits a loss of more than $3$ dB.

\subsection{Coded MIMO Systems}

\begin{figure}
\centerline{\psfig{figure=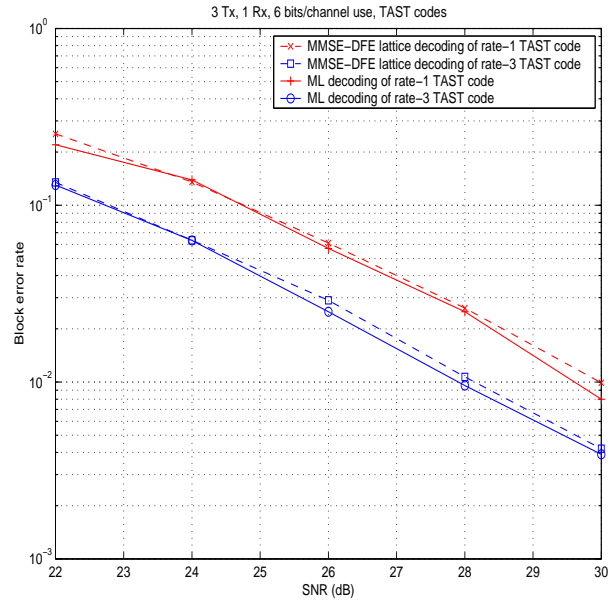,width=8cm,height=8cm}}
\caption{Performance of TAST codes under MMSE-DFE lattice decoding
  and ML detection with $M=3$ and $N=1$.}
              \label{figure0}
\end{figure}

In this section, we consider two classes of space-time codes. The
first class is the linear dispersion (LD) codes which are obtained
by applying a linear transformation (over ${\mathbb C}$) to a
vector of PAM symbols. For convenience, we follow the set-up of
Dayal and Varanasi \cite{Mahesh} where two variants of the
threaded algebraic space-time (TAST) constellations \cite{tast}
are used in a $3\times 1$ MIMO channel. This setup also allows for
demonstrating the efficiency of the MMSE-DFE frontend in solving
under-determined systems. In \cite{Mahesh}, the rate-$1$ TAST
constellation uses $64$-QAM inputs at a rate of one symbol per
channel use. The rate-$3$ TAST constellation, on the other hand,
uses $4$-QAM inputs to obtain the same throughput as the rate-$1$
constellation. As observed in \cite{Mahesh}, one obtains a sizable
performance gain when using rate-$3$ TAST constellation under ML
decoding. The main disadvantage, however, of the rate-3 code is
that it corresponds to an under-determined system with $6$ excess
unknowns which significantly complicates the decoding problem.
Fig.~\ref{figure0} shows that the performance of the proposed
MMSE-DFE lattice decoder is less than $0.1$ dB away from the ML
decoder for both cases. In order to quantify the complexity
reduction offered by our approach, compared with the generalized
sphere decoder (GSD) used in \cite{Mahesh}, we measure the average
complexity increase with the excess dimensions. If we define
\begin{equation} \label{bench}
\gamma \defines \frac{\mbox{Average complexity of decoding
rate-$3$ constellation}}{\mbox{Average complexity of decoding
rate-$1$ constellation}},
\end{equation}
then a straightforward implementation of the GSD, as outlined in
\cite{gsd} for example, would result in $\gamma={\cal
O}\left(4^6\right)$. In fact, even with the modification proposed
in \cite{Mahesh}, Dayal and Varanasi could only bring this number
down to $\gamma=460$ at an SNR of $30$ dB. In Table~\ref{table1},
we report $\gamma$ for the proposed algorithm at different SNRs,
where one can see the significant reduction in complexity (i.e.,
from $460$ to $12$ at an SNR of $30$ dB). Based on experimental
observations, we also expect this gain in complexity reduction to
increase with the excess dimension $m-n$.

\begin{figure}   
\centerline{\subfigure{\epsfig{file=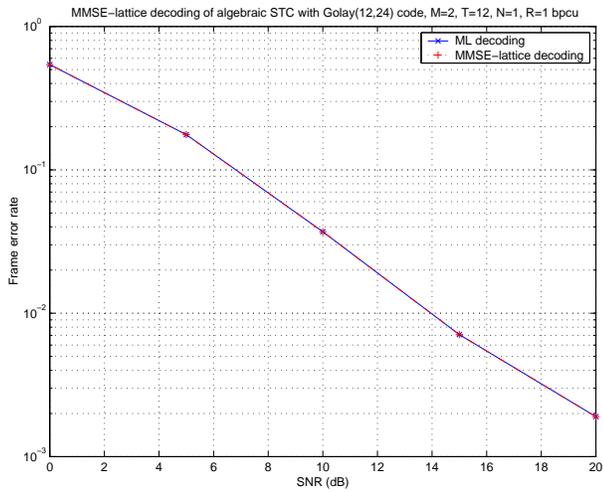,width=8.0cm}}
\hfill \subfigure{\epsfig{file=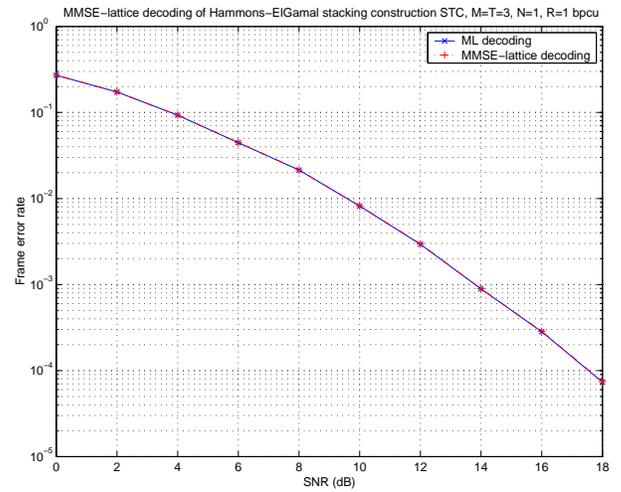,width=8.0cm}}}
\caption{Performance of MMSE-DFE lattice decoding and ML decoding
for algebraic space-time codes} \label{fig_stcode}
\end{figure}

The second space-time coding class is the algebraic codes proposed
in \cite{RHHG,LK,LFT}. This approach constructs linear codes, over
the appropriate finite domain, and then the encoded symbols are
mapped into QAM constellations. The QAM symbols are then parsed
and appropriately distributed across the transmit antennas to
obtain full diversity. It has been shown that the complexity of ML
decoding of this class of codes grows exponentially with the
number of transmit antennas and data rates. Here, we show that the
proposed tree search framework allows for an efficient solution to
this problem. Figure~\ref{fig_stcode} shows the performance of MMSE-DFE
lattice decoding for two such constructions of space-time codes {\it i.e.,}
Golay space-time code for two transmit antennas and the companion matrix
code for three transmit antennas \cite{RHHG}.
 In both
case, the performance of the MMSE-DFE lattice decoder is seen to be essentially same
as the ML performance. In the proposed decoder, we use the lattice
${\Lambdam}$ obtained from underlying algebraic code through
construction~A. The ML performance, obtained via exhaustive search in
Figure~\ref{fig_stcode}, is not feasible for higher dimensions due to
exponential complexity in the number of dimensions.

\subsection{Coded Transmission over ISI Channels}
\begin{figure}   
\centerline{\subfigure{\epsfig{file=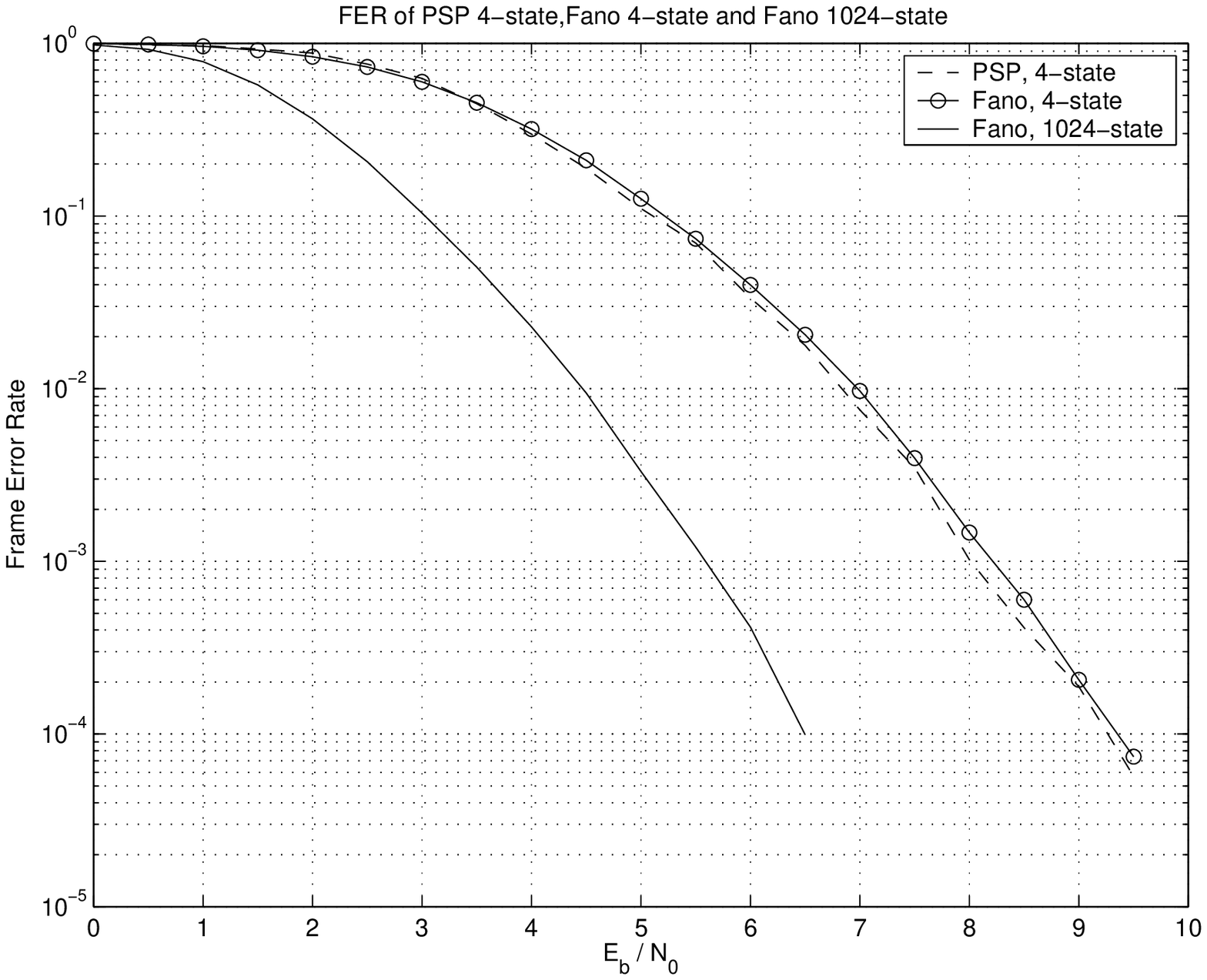,width=8.0cm}}
\hfill \subfigure{\epsfig{file=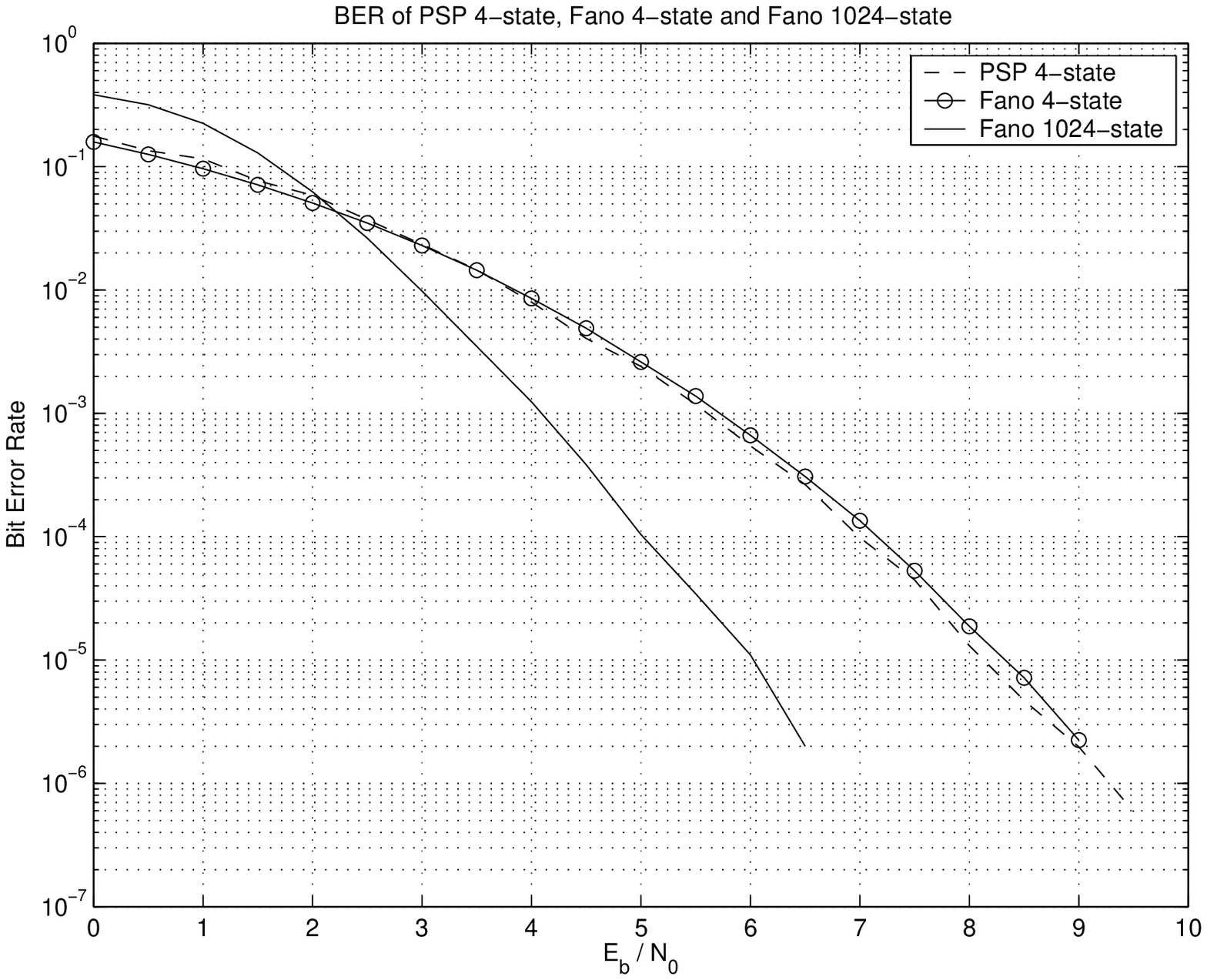,width=8.0cm}}}
\caption{Frame and Bit Error Rate curves for the Fano and PSP algorithms
for convolutional codes over an ISI channel} \label{fig_isi}
\end{figure}

In this section, we compare the performance of the MMSE-Fano decoder
with the Per-Survivor-Processing (PSP) algorithm for convolutionally coded
transmission over ISI channels. Our MMSE-Fano decoder uses the
construction~A lattice obtained from the convolutional code.
For this scenario, it is known 
that PSP achieves near-ML frame error rate performance \cite{caire_psp}.
Figure~\ref{fig_isi} compares the Frame and
Bit Error Rates for a $4-$state, rate $1/2$ convolutional code with generator
polynomials given by $(5,7)$ and code length $200$, over a $5-$tap ISI
channel. The channel impulse response was chosen as
$(0.848,-0.424,0.2545,-0.1696,0.0848)$. The Fano decoder with $b=1$ and
stepsize $1$ is seen to achieve
essentially the same performance as the PSP algorithm for this code, with
reasonable search complexity over the entire SNR range. We again note that
the loss in lattice decoding as opposed to finite search space is
negligible, due to MMSE-DFE preprocessing of the channel prior to the search.
Moreover, the complexity of PSP algorithm, although linear in frame length,
increases exponentially with the constraint length of the convolutional
code used, while that of the Fano decoder is essentially independent of the
constraint length. Figure~\ref{fig_isi} also shows the performance of the
Fano decoder for a rate $1/2$, $1024$-state convolutional code with
generator polynomials $(4672,7542)$, with the same frame
size. Due to the increased constraint length, the performance is
significantly better (with almost no increase in complexity). The
complexity of PSP algorithm, on the other hand, is significantly higher for
this code.

%---------------------------------------------------------------------------------
%---------------------------------------------------------------------------------
\section{Conclusions} \label{conclusions}
A central goal of this paper was to introduce a unified framework
for tree search decoding in wireless communication applications.
Towards this end, we identified the roles of two different, but
inter-related, components of the decoder, namely; 1) Preprocessing
and 2) Tree Search. We presented a preprocessing stage composed of
MMSE-DFE filtering for left preprocessing and lattice reduction
with column ordering for right preprocessing. We argued that this
preprocessor allows for ignoring the boundary control in the tree
search stage while entailing only a marginal loss in performance.
By relaxing the boundary control, we were able to build a generic
framework for designing tree search strategies for joint detection
and decoding. Within this framework, BeFS emerged as a very
efficient solution that offers many valuable advantages. To limit
the storage requirement of BeFS, we re-discovered the Fano decoder
as our proposed tree search algorithm. Finally, we established the
superior performance-complexity tradeoff of the Fano decoder
analytically in a V-BLAST configuration and demonstrated its
excellent performance and complexity in more general scenarios via
simulation results.
\appendix

%-----------------------------------------------------------------------------------

%---------------------------------------------------------------------------------------------

\section{The Fano Decoder} \label{app_fano_algo}

In this section, we obtain the cost function used in the proposed
Fano/Stack decoder from the Fano metric defined for tree codes
over general point-to-point channels, and give a brief description
of the Fano decoder and its properties.

\subsection{Generic Cost Function of the Fano Decoder}
For the transmitted sequence $\hat{\mb x}$, let
\begin{equation} \label{fano_metric:1}
\mathbf{y}=\mathbf{R}\hat{\mb x}+\mathbf{w}
\end{equation}
be the system model, as in Section~\ref{system}. In
(\ref{fano_metric:1}), the noise sequence ${\mb w}$ is composed of
{\it i.i.d} Gaussian noise components with zero mean and unit
variance.

For a general point-to-point channel with continuous output, the
Fano metric of the node ${\mathbf x}_1^k$ can be written as
\cite{johann}
\begin{equation}
\label{metric0}
\mu(\mathbf{x}_1^k)=\log\left(\frac{Pr(\mathcal{H}(\mathbf{x}_1^k))p(\mathbf{y}_1^k|\mathcal{H}(\mathbf{x}_1^k))}{p(\mathbf{y}_1^k)}\right)
\end{equation}
where $\mathcal{H}({\mathbf x}_1^k)$ is the hypothesis that
${\mathbf x}_1^k$ form the first $k$ symbols of the transmitted
sequence.

For $1\leq k\leq m$, if $Pr(\mathcal{H}(\mathbf{x}_1^k))$ is
uniform over all nodes $\mb{x}_1^k$ that consist of the first $k$
components of any valid codeword in ${\mathcal C}$, from
(\ref{metric0}), the cost function for the Fano decoder for our
system model (\ref{fano_metric:1}) can be simplified as

\begin{equation} \label{fano_metric:2}
f({\mb x}_1^k)=-\mu({\mb x}_1^k)=\log\left(\sum_{\mathbf{x}_1^k}
e^{-\frac{\sum_{j=1}^k
w_j(\mathbf{x}_1^j)}{2}}\right)+\frac{\sum_{j=1}^k
w_j(\mathbf{x}_1^j)}{2}.
\end{equation}
Since summation over $\mb{x}_1^k$ in (\ref{fano_metric:2}) is not
feasible, we use the following approximations: first, $\log(\sum
a_i)\approx \log(\max(a_i))$, so the sum can be approximated by
the largest term. Second, for moderate to high SNRs, the
transmitted sequence is actually the closest vector with a high
probability, {\it i.e.,} the largest term corresponds to the
transmitted sequence. Thus, (\ref{fano_metric:2}) can be
approximated as
\begin{equation}
\label{fano_metric:3} \log\left(\sum_{\mathbf{x}_1^k}
e^{-\frac{\sum_{j=1}^k w_j(\mathbf{x}_1^j)}{2}}\right) \approx
-\frac{|\mb{w}_1^k|^2}{2}.
\end{equation}
After averaging (\ref{fano_metric:3}) over noise samples and
scaling, we have,
\[f({\mb x}_1^k)=\sum_{j=1}^k w_j({\mb x}_1^j)-k\]

In general, the cost function for the Fano/Stack decoder can be
written in terms of the parameter $b$, the bias, as
\[
f({\mb x}_1^k)=\sum_{j=1}^k w_j({\mb x}_1^j)-bk.
\]

\subsection{The Algorithm}
The operation of the Fano decoder with no boundary control
(lattice decoding) follows the following steps:

\begin{itemize}

\item {\it Step 1: (Initialize)} Set $k\leftarrow 0$, $T\leftarrow
0$,
  $\mb{x}\leftarrow x_0$.

\item {\it Step 2: (Look forward)} ${\mb x}_1^{k+1} \leftarrow
({\mb x}_1^k,x_{k+1})$, where $x_{k+1}$ is the $(k+1)^{th}$
component of the best child node of ${\mb x}_1^k$.

\item {\it Step 3:}

If $f(\mb{x}_1^{k+1})\leq T$,

\hspace{0.1in} If $k+1=m$ {\it (leaf node)}, then $\hat{\mb
x}={\mb x}_1^m$; exit.

\hspace{0.1in} Else {\it (move forward)}, $k\leftarrow k+1$.

\hspace{0.2in} If $f(\mb{x}_1^{k-1})>T-\Delta$,

\hspace{0.3in} while $f(\mb{x}_1^{k})\leq T-\Delta$, $T\leftarrow
T-\Delta$ {\it (tighten threshold)}.

\hspace{0.2in} {\it Go to} step 2.

Else

\hspace{0.1in} If $(k=0$ or $f(\mb{x}_1^{k-1})>T)$, $T\leftarrow
T+\Delta$ {\it (cannot move back, so relax threshold}).

\hspace{0.2in} {\it Go to} step 2.

\hspace{0.1in} Else {\it (move back and look forward to the next
best node)}

\hspace{0.2in} ${\mb x}_1^k\leftarrow \{{\mb x}_1^{k-1},x_k\}$,
where $x_k$ is the last component of the next best child node of
${\mb x}_1^{k-1}$.

\hspace{0.2in} $k\leftarrow k-1$.

\hspace{0.2in} {\it Go to} Step 3. \hfill $\Box$

\end{itemize}

Note that $T$ (i.e., the threshold) is allowed to take values only
in multiples of the step size $\Delta$ (i.e., $0,\pm \Delta,\pm
2\Delta,...$).
When a node is visited by the Fano decoder for the
first time, the threshold $T$ is tightened to the least possible
value while maintaining the validity of the node. If the current
node does not have a valid child node, then the decoder moves back to the
parent node (if the parent node is valid) and attempts moving forward to
the next best node. However, if the
parent node is not valid, the threshold is relaxed
and attempt is made to move forward again, proceeding in this way
until a leaf node is reached.

The determination of best and next best child nodes is simplified
in CLPS problem; the child node generation order {\it gen} in SE
enumeration (section~\ref{dfs}) generates child nodes with cost
functions in ascending order, given any node ${\mb x}_1^k$.

\subsection{Properties of the Fano Decoder} \label{basic_lemmas}

The main properties of the Fano decoder used in our analysis are
\cite{johann}:

\begin{enumerate}
\item A node ${\mathbf x}_1^k$ is generated by the Fano decoder
only if its cost function is not greater than the bound $T$.

\item Let {\it correct path} be defined as the path corresponding
to the transmitted codeword, and let $f_{M}$ be the maximum cost
function along the correct path.  The bound $T$ is always less
than $(f_{M}+\Delta)$, where $\Delta$ is the step-size of the Fano
decoder; that is, $\max\{T\} < T_M \defines f_M+\Delta $.

\end{enumerate}

All nodes that are generated by the Fano decoder are necessarily
  those with
cost function less than the bound $T$, by Property~$(1)$. However,
even though the cost function of some node ${\mb x}_1^k$ may be
smaller than the bound, the node itself might not be visited when
bound takes the value $T$. If any of the cost functions along the
path $\{{\mb x}_1^r, r<k\}$ increases above $T$, the node ${\mb
x}_1^r$ is not generated and thus ${\mb x}_1^k$ is not visited.
Hence, this is not a sufficient condition for a node to be
generated.

Moreover, in Property~$(2)$, the bound $T$ is always lesser than
$(f'_{M}+\Delta)$, where $f'_{M}$ is the maximum cost function
along any path of length $m$. A tight bound is obtained only when
the maximum cost function corresponding to the path with the least
$f'_{M}$ is chosen. However, $f_{M}$ along the transmitted path is
usually easier to characterize statistically than $f'_{M}$.

%--------------------------------------------------------------------------

\section{Properties of the Stack Decoder} \label{stack_prp}

For any node $\xv_1^k$ in the tree, let $h(\xv_1^k)\defines \sum_{i=1}^k
w_i(\xv_1^k)-bk$. For the stack algorithm, the cost function of any node in
ACTIVE at any instant defined as follows: If $\xv_1^k$ is a leaf node, then
$f(\xv_1^k)=-\infty$. Otherwise, we let $\xv_{1,g}^{k+1}$ be the best child node of
$\xv_1^k$ not generated yet, and define $f(\xv_1^k)=h(\xv_{1,g}^k)$. We note that
$h$ of any node, once generated, remains constant throughout the algorithm,
and $f$ of any node is non-decreasing as the algorithm progresses.

\begin{proposition} 
Let $\bar{\xv}_1^m=(\bar{x}_1,...\bar{x}_m)$ be the
path chosen by the stack algorithm, and $\xv_1^m=(x_1,...,x_m)$ be
any path in the tree. Then, 
\begin{equation} \label{stack_prp1}
\max_{1\leq j\leq m} h(\bar{\xv}_1^j)\leq \max_{1\leq j\leq m} h(\xv_1^j)
\end{equation}
\end{proposition}

\begin{proof}
On the contrary, assume there exists a path
$(\bar{x}_1,\bar{x}_2,...,\bar{x}_d,\breve{x}_{d+1},...,\breve{x}_m)$
that does not satisfy (\ref{stack_prp1}). Here, the
path is assumed to share the same nodes with the chosen path until level
$d$, and diverges from the chosen path from level $d+1$ onwards. Since this
path does not satisfy (\ref{stack_prp1}),
\begin{equation} \label{stk_p2}
\max_{d+1\leq j\leq m} h(\bar{\xv}_1^j) > \max_{d+1\leq j\leq m} h(\breve{\xv}_1^j)
\end{equation}
Let $\bar{\xv}_1^k$, $k>d$, be the node for which $\max_{d+1\leq j\leq m} h(\bar{\xv}_1^j)$ occurs. 
Then, we have,
\begin{equation} \label{stk_p3}
h(\bar{\xv}_1^k) > h(\breve{\xv}_1^j), \quad d<j\leq m
\end{equation}
Since  $\bar{\xv}_1^m$ is the chosen path, the node
$\bar{\xv}_1^k$ is generated at some instant before the search
terminates. Just before $\bar{\xv}_1^k$ is generated, $h(\bar{\xv}_1^{k-1})=g(\bar{\xv}_1^k)$, since
$\bar{\xv}_1^k$ is the best child node of $\bar{\xv}_1^{k-1}$ not generated
yet. Moreover, since $h(\bar{\xv}_1^k)>h(\breve{\xv}_1^{d+1})$, the node
$\bar{\xv}_1^d$ with cost function
$f(\bar{\xv}_1^d)=h(\breve{\xv}_1^{d+1})$ appears at the top of the stack
at some instant before $\bar{\xv}_1^k$ is generated. Therefore,
$\breve{\xv}_1^{d+1}$ is generated before $\bar{\xv}_1^k$ is
generated. Since the search does not terminate before $\bar{\xv}_1^k$ is
generated, applying the same argument, one sees that all the nodes
$\breve{\xv}_1^{d+2},...,\breve{\xv}_1^{m}$ are generated before
$\bar{\xv}_1^k$ is generated. However, once $\breve{\xv}_1^{m}$ is
generated by the stack algorithm, the search terminates, with
$(\bar{x}_1,...,\bar{x}_d,\breve{x}_{d+1},...,\breve{x}_m)$
as the chosen path. Since $1\leq d\leq m$ can take any value, the
inequality in (\ref{stack_prp1}) is satisfied by all paths.
\end{proof}

\begin{proposition}
If 
\begin{equation} \label{stk_p4}
\max_{1\leq j\leq d} h(\xv_1^d) > \max_{1\leq j\leq m} h(\bar{\xv}_1^j),
\end{equation}
then, the node $\xv_1^d$ is not generated.
\end{proposition}

\begin{proof}
First, we show that if 
\begin{equation} \label{stk_p5}
h(\xv_1^d)> \max_{1\leq j\leq m} h(\bar{\xv}_1^j),
\end{equation}
then $\xv_1^d$ is not generated.
Let (\ref{stk_p5}) be true, and assume $\xv_1^d$ is generated. Then, just before $\xv_1^d$
is generated, its parent node $\xv_1^{d-1}$ is at the top of ACTIVE, with
cost function $f(\xv_1^{d-1})=h(\xv_1^d)$. However, since $h(\xv_1^d) >
h(\bar{\xv}_1^j),\quad 1\leq j\leq m$, all nodes along the chosen path are
generated before $\xv_1^d$ is generated, and the hence the search
terminates before $\xv_1^d$ is generated. 
Noting that $\xv_1^d$ can be generated only if all the nodes
$\xv_1^1,...,\xv_1^{d-1}$ are generated, and applying the same argument for
$\xv_1^{d-1},...,\xv_1^1$, we
have (\ref{stk_p4}).
\end{proof}

%--------------------------------------------------------------------------

\section{Proof of Theorem~\ref{thm_bfs2}} \label{bfs_vs_dfs}

%Let ${\mathcal B_{SE}}$ and ${\mathcal B_{IR}}$ correspond to the
%set of nodes generated by SE enumeration and the IR algorithm,
%respectively. Also, let ${\mathcal B_{s,0}}$ and ${\mathcal
%B_{s,b}}$ correspond to the set of nodes generated by the stack
%decoder with bias $b=0$ and with bias $b>0$, respectively.

Let ${\mathcal A_{IR}}$ be the
set of nodes generated by the IR algorithm, where the bounding function $\tv$ has
components given by $t_k=bk+\delta$. Let ${\mathcal A_s}$ be the set of nodes generated by the stack
decoder with the bias $b$.
The IR algorithm in Theorem~\ref{thm_bfs2} can be defined with
bounding function given by $\{t_k=bk+\delta,\quad 1\leq k\leq m\}$, and
the cost function for any node ${\mb x}_1^k$ given by
$\sum_{i=1}^k w_i({\mb x}_1^i)$, or equivalently, with the
bounding function $t_k=\delta$ and cost function $\left(\sum_{i=1}^k w_i({\mb x}_1^i)-bk\right)$. If
$\delta$ is the bound of
the IR algorithm, then any node ${\mathbf x}_1^k$ is generated by
the algorithm if and only if all the conditions 
$\left\{\sum_{i=1}^1 w_i({\mb x}_1^i)-b<\delta,\sum_{i=1}^2 w_i({\mb x}_1^i)-2b<\delta,...,\sum_{i=1}^k w_i({\mb x}_1^i)-bk<\delta\right\}$,
%$\{f({\mathbf
%x}_1^1)<\delta,f({\mathbf x}_1^2)<\delta,...,f({\mathbf x}_1^k)<\delta\}$ are
are satisfied. Therefore,
\begin{equation}
\label{sp} \mathcal{A}_{IR}=\left\{ \mathbf{x}_1^k : \max_{1\leq j\leq
k} \left(\sum_{i=1}^k w_i({\mb x}_1^i)-bk\right) < \delta\right\}.
\end{equation}
Moreover, $\delta$ should be such that at least one sequence  ${\mathbf
x}\in \Uc$ is included within the search space.\footnote{Otherwise,
$\delta$ is increased and search is repeated afresh} Let $\hat{\mathbf
x}_{IR}$ be a leaf node such that
\begin{equation} \label{ir1}
\hat{\mathbf x}_{IR}=\arg \min_{\mathbf{x}\in \Uc} \left(\max_{1\leq
k\leq m} \left(w_i({\mb x}_1^i)-bk\right)\right).
\end{equation}
{\it i.e.,} $\hat{\mathbf x}_{IR}$ has the least value of maximum
cost function among all paths of length $m$. If 
\[
\delta<\max_{1\leq j\leq m} \left(w_i(\hat{\mathbf x}_{IR,1}^j)-bk\right),
\]
then no ${\mathbf x}\in \Uc$ lies within the search space, and the search space is empty.
If lattice decoding is used, then the minimum in (\ref{ir1}) is taken over all
$\mathbf{x}\in {\mathbb Z}^m$.
Therefore, $\delta > \max_{1\leq j\leq m} \left(w_i(\hat{\mathbf x}_{IR,1}^j)-bk\right)$.
From Section~\ref{stack_prp}, Prop.~1, the path chosen by the stack
algorithm, $\xv_1^m$ satisfies
\begin{equation} \label{stack_prop1}
\left(\max_{1\leq k\leq m} \left(w_i(\bar{\mb x}_1^i)-bk\right)\right)\leq \left(\max_{1\leq k\leq m} \left(w_i({\mb x}_1^i)-bk\right)\right)
\end{equation}
where $\xv_1^m$ is any other path.

From (\ref{ir1}) and (\ref{stack_prop1}), 
\begin{equation} \label{stack_prop2}
\left(\max_{1\leq k\leq m} \left(w_i(\bar{\mb
x}_1^i)-bk\right)\right)=\left(\max_{1\leq k\leq m} \left(w_i(\hat{\mb
x}_{IR,1}^i)-bk\right)\right)<\delta
\end{equation}
From Proposition~2 and (\ref{stack_prop1}), ${\Ac_s}\subseteq {\Ac_{IR}}$.

%-----------------------------------------------------------------

\section{Proof of Theorem~\ref{thm_diversity}} \label{proof_of_diversity}

In this section, we derive an upper bound to the frame error rate
for a V-BLAST system with uncoded input (with $Q$-PAM
constellation for the components), for the Fano decoder that
visits paths in the regular $Q$-PAM signal space. The
preprocessing assumed here is QR transformation of ${\mathbf H}$.

Let ${\mathcal E}_f$ be the event that the Fano decoder makes an
erroneous detection, conditioned on $T_M \defines f_M+\Delta $.
Then, $P_e=E_{T_{M}}(Pr({\mathcal E}_f))$ is the frame error rate
of the Fano decoder. In this section, we derive an upper bound on
$P_e$.
From property $(2)$ in Section~\ref{basic_lemmas}, $T < (f_M+\Delta)$, where $\Delta$ is the step size of the
Fano decoder. Any sequence ${\mb x}\neq {\hat {\mb x}}$ can be
decoded as the closest point by the Fano decoder only if its cost
function is lesser than $T_{M}$. One has
\begin{eqnarray}
\mb{y}&=&\mb{H}\mb{x}+\mb{z}
=\mb{Q}\begin{pmatrix} \Rm \\ \mb{0} \end{pmatrix}\mb{x}+\mb{z},
\end{eqnarray}
and therefore
\begin{eqnarray}
\yv &\leftarrow& \mb{Q}^\transp\yv=\begin{pmatrix} \Rm \\ \mb{0}
\end{pmatrix}\mb{x}+\begin{pmatrix} \mb{w}_{r+1}^n \\ \mb{w}_1^{r} \end{pmatrix}
\end{eqnarray}
where $r=n-m$ is the excess degrees of freedom in the V-BLAST
system. Since the cost function of a leaf node ${\mb x}_1^m$ is
$\displaystyle{f({\mb x}_1^m)=\sum_{i=1}^m w_i({\mb
x}_1^i)-bm=|\mb{R}\tilde{\mb x}+\mb{w}_{r+1}^n|^2}-bm$, $P({\mathcal
E}_f)$ can be upper bounded as
\begin{eqnarray}
\label{fer1}
P(\mc{E}_f)&\leq&\sum_{{\mb x}\in \Uc,{\mb x}\neq\hat{\mb x}}
\Pr(\sum_{j=1}^m w_j(\mb{x}_1^j)-bm<T_{M}) \\
\label{wer1} &=&\sum_{{\mb x}\in \Uc,{\mb x}\neq\hat{\mb x}}
\Pr(|\mb{R}\tilde{\mb{x}}+\mb{w}_{r+1}^n|^2<bm+f_{M}+\Delta)
\end{eqnarray}
where
\[
f_M=\max\left\{0,|\mb{w}_{r+1}^{r+1}|^2-b,|\mb{w}_{r+1}^{r+2}|^2-2b,
  \hdots,|\mb{w}_{r+1}^{n}|^2-mb\right\}
\]
is the maximum cost function along the transmitted sequence path.
The upper bound in (\ref{fer1}) follows from the union bound, and due to
the fact that in general, $f({\mb x}_1^m)<T_M$ is only a necessary
condition for ${\mb x}_1^m$ to be decoded by the Fano decoder. 

The bound in (\ref{wer1}) can be rewritten as
\begin{eqnarray}
P(\mc{E}_f)&\leq&\sum_{{\mb x}\in \Uc,{\mb x}\neq\hat{\mb
x}}\Pr\left(\left|\bmx \mb{R}\\ \mb{0}\emx
\tilde{\mb{x}}+\mb{w}_{1}^n\right|^2<bm+f_{M}+\Delta+|\mb{w}_1^r|^2\right)\\
&=&\sum_{{\mb x}\in \Uc,{\mb x}\neq\hat{\mb x}}\Pr\left(\left|\bmx \mb{R}\\
\mb{0}\emx\tilde{\mb{x}}+\mb{w}_{1}^n\right|^2-|\mb{w}_{1}^n|^2<bm+f_{M}+\Delta-|\mb{w}_{r+1}^n|^2\right)\\
&=&\sum_{{\mb x}\in \Uc,{\mb x}\neq\hat{\mb
x}}\Pr(|\mb{H}\tilde{\mb{x}}|^2+2(\mb{H}\tilde{\mb{x}})^\transp\mb{z}<bm+f_{M}+\Delta-|\mb{w}_{r+1}^n|^2)\\
\label{div:1} &\leq&\sum_{{\mb x}\in \Uc,{\mb x}\neq\hat{\mb
x}}\Pr(|\mb{H}\tilde{\mb{x}}|^2+2(\mb{H}\tilde{\mb{x}})^\transp\mb{z}<bm+\Delta),
\end{eqnarray}
since
$f_{M}-|\mb{w}_{r+1}^n|^2=\max\left\{-|\mb{w}_{r+1}^n|^2,-|\mb{w}_{r+2}^n|^2-b,...,-mb\right\}\leq
0$. The bound in (\ref{div:1}) is now independent of the value of
$f_M$, and hence represents a bound on the frame error rate.
Note that the corresponding expression in (\ref{div:1}) for ML
decoding is
$\Pr(|\mb{H}\tilde{\mb{x}}|^2+2(\mb{H}\tilde{\mb{x}})^\transp\mb{z}<0)$.
For any ${\mb x}\in \Uc$ and ${\mb x}\neq \hat{\mb x}$, let
$d^2({\hat{\mb x}},\mb{x})=|\mathbf{H}\tilde{\mb x}|^2$ represent
the squared Euclidean distance between the lattice points ${\mb
H}{\mb x}$ and ${\mb H}\hat{\mb x}$. Then,
\begin{equation}
\label{wer4a} \Pr(|{\mb H}\tilde{\mb{x}}+\mb{z}|^2-|\mb{z}|^2\leq
mb+\Delta)\leq
\begin{cases}
e^{\left(-\frac{1}{8}{(d^2({\hat{\mb
x}},\mb{x})-mb-\Delta)^2}/{d^2({\hat{\mb x}},\mb{x})}\right)},&
d^2({\hat{\mb
x}},\mb{x})>mb+\Delta \\
1 &  d^2({\hat{\mb x}},\mb{x})\leq mb+\Delta \\
\end{cases}
\end{equation}
by Chernoff bound.

For $d^2({\hat{\mb x}},\mb{x})>mb+\Delta$, equation (\ref{wer4a})
can be rewritten as
\begin{eqnarray}
\Pr(|{\mb H}\tilde{\mb{x}}+\mb{z}|^2-|\zv|^2\leq mb+\Delta)&\leq&
e^{-\frac{1}{8}\left(d^2(\hat{\mb{x}},\mb{x})+\frac{(mb+\Delta)^2}{d^2(\hat{\mb{x}},\mb{x})}-2(mb+\Delta)\right)}
\\
\label{wer4b} &\leq&
e^{-\frac{1}{8}\left(d^2(\hat{\mb{x}},\mb{x})\right)}e^{(mb+\Delta)/4}
\end{eqnarray}
since $e^{-(mb+\Delta)^2/(8d^2(\hat{\mb{x}},\mb{x}))}<1$, for
$d^2(\hat{\mb{x}},\mb{x})>0$. Let

\begin{eqnarray}
q&\defines& \min_{\mb{x}_i,\mb{x}_j\in \Uc,i\neq
  j}(|\mb{H}(\mb{x}_i-\mb{x}_j)|^2)
\end{eqnarray}
and let
\begin{eqnarray}\label{pegq}
g(q)&\defines&
\begin{cases}
\sum_{{\mb x}\in \Uc,\tilde{\mb x}\neq{\mb
    0}}e^{(mb+\Delta)/4} e^{\left(-q/8\right)}& q>mb+\Delta \\
1 &  q<mb+\Delta \\
\end{cases}
\end{eqnarray}
Then, from (\ref{wer4b}) and (\ref{wer4a}), $P_e\leq E_q(g(q))$.
An upper bound on the probability density function (pdf) of $q$ is
given by \cite{nee}
\begin{eqnarray}\label{pegq2}
p(q)<p_{\chi}(q)\sum_{k=1}^m \left(\begin{array}{c} m \\ k \\
  \end{array}\right)\frac{1}{k}
\end{eqnarray}
where $p_{\chi}(q)$ is the pdf of a scaled chi-square random
variable with $n$ degrees of freedom and mean $\frac{n\rho}{m}$
(i.e., a random variable that is the sum of squares of $n$ i.i.d
zero-mean Gaussian variables with variance $\frac{\rho}{m}$).
Then, \eqref{pegq} and \eqref{pegq2} give
\begin{eqnarray}
P_e&\leq& Q^m e^{(mb+\Delta)/4} \int_{mb+\Delta}^{\infty}
e^{\left(-q/8\right)} p(q) dq + \int_{0}^{mb+\Delta} p(q) dq\\
&=& AQ^m e^{(mb+\Delta)/4} \int_{mb+\Delta}^{\infty}
e^{\left(-q/8\right)} \frac{q^{(n/2-1)}e^{-q/(2\sigma^2)}}{2^{n/2}
\Gamma(\frac{n}{2})
  \sigma^n} dq
+A \gamma\left(\frac{mb+\Delta}{2\sigma^2},\frac{n}{2}\right) \\
&\leq& AQ^m e^{(mb+\Delta)/4} \int_{0}^{\infty}
e^{\left(-q/8\right)} \frac{q^{(n/2-1)}e^{-q/(2\sigma^2)}}{2^{n/2}
\Gamma(\frac{n}{2})
  \sigma^n} dq
+A \gamma\left(\frac{mb+\Delta}{2\sigma^2},\frac{n}{2}\right) \\
&=& \frac{AQ^m
  e^{(mb+\Delta)/4}}{\left(1+\frac{\rho}{4m}\right)^{n/2}}+A
\gamma\left(\frac{mb+\Delta}{2\sigma^2},\frac{n}{2}\right)
\end{eqnarray}
where $\sigma^2=\frac{\rho}{m}$,
$A\defines \displaystyle \sum_{k=1}^m  \frac{1}{k} {m \choose k}$
is a constant independent of $q$ or $\rho$, and $\gamma(x,a)$
is the incomplete gamma function. If $b$ is bounded (i.e.,
$b<{\mathcal M}<\infty$) $\forall \rho$, then $e^{(mb/4)}$ is also
bounded for all $\rho$ and finite $m$. The error performance of
the Fano decoder can now be characterized by the sum of two terms.
The dependence of the first error term on $\rho$ is of the form
$\rho^{-(n/2)}$ for large values of SNR, and hence has the same
diversity as the ML decoder. The second term can also be bounded
as
\begin{eqnarray}
\label{st1}
\gamma\left(\frac{mb+\Delta}{2\sigma^2},\frac{n}{2}\right)&\leq&
\left(1-e^{-(mb+\Delta)/(2\sigma^2)}\right)^{(n/2)} \\
\label{st2}
&\leq& \left(\frac{mb+\Delta}{2\sigma^2}\right)^{(n/2)} \\
&=& \left(\frac{m(mb+\Delta)}{2\rho}\right)^{(n/2)}
\end{eqnarray}
where (\ref{st1}) follows from the inequality $\gamma(x,a)\leq
(1-e^{-x})^a$ (Appendix~\ref{inc_gamma_ub}), and (\ref{st2}) from $(1-e^{-x})<x$ for $x>0$. The
second term also has the dependence $\rho^{-(n/2)}$, and hence the
Fano decoder achieves the {\it same} diversity as that of the ML
decoder for this system.

The above derivation also applies to the Stack algorithm, with
minor modifications. Let ${\mathcal E}_s$ be the event that the
stack algorithm makes an erroneous detection, conditioned on the
value of $f_M$. Then, $P_e=E_{f_M}(\Pr({\mathcal E}_s))$ is the
word error rate of the stack algorithm.
Since any path ${\mb x}\neq {\hat {\mb x}}$ is decoded as the
closest point by the stack algorithm only if $h({\mb x})=\sum_{i=1}^m w_i({\xv}_1^i)-bm$ is
not greater than $f_M$ (Prop.~1, Section~\ref{stack_prp} , $P({\mathcal E}_s)$ can be written as
\begin{eqnarray}
P(\mc{E}_s)&\leq& \sum_{{\mb x}\in \Uc,{\mb x}\neq\hat{\mb
    x}} \Pr\left\{\sum_{j=1}^m w_j(\mb{x}_1^j)-bm<f_M\right\} \\
\label{wer1stack} &=& \sum_{{\mb x}\in \Uc,{\mb x}\neq\hat{\mb x}}
\Pr\left\{|{\mb H}\tilde{\mb{x}}+\mb{z}|^2<bm+f_M\right\}
\end{eqnarray}
From (\ref{wer1stack}) and (\ref{wer1}), it is easy to see that the
error probability expression for the stack algorithm is the same
as that for the Fano decoder, when ${\Delta}=0$. Thus, the stack
algorithm too achieves the same diversity as the ML decoder for a
V-BLAST system, for any finite value of $b$.  \hfill $\Box$

\section{Proof of Theorem~\ref{thm_stack_comp_2}} \label{stack_complexity}
%-------------------------------------------------------------------------------
The following are required for the proof.

\subsection{Wald's inequality} \label{wald_ineq}

Let $S_0=0,S_1,S_2,...$ be a random walk, with $S_j=\sum_{i=1}^{j}
X_i$, where $X_i$s are {\it i.i.d} random variables such that
$\Pr(X_i>0)>0$, $\Pr(X_i<0)>0$, and $E(X_i)<0$. Let
$g(\lambda)=E(e^{\lambda X_i})$ be the moment generating function
of $X_i$.
Let $\lambda_0 > 0$ be a root of $g(\lambda)=1$. Then, from Wald's
identity \cite{johann},
\begin{equation}
\label{umin1} \Pr(S_{\max}>u)\leq e^{-\lambda_0 u}
\end{equation}
where $S_{\max}=\max_j (S_j)$.

For the random walk with $X_i=w_i^2-b$, where $w_i\sim {\mathcal
N}(0,1)$, the above conditions are satisfied if $b>1$. The moment
generating function for $X_i=w_i^2-b$ is given by
\begin{equation} \label{mgf}
g(\lambda)=\frac{e^{-\lambda b}}{\sqrt{1-2\lambda}}.
\end{equation}
From (\ref{mgf}), $\lambda_0>0$ can be found as the positive root of the
equation
\[
-2\lambda b = \log(1-2\lambda).
\]
Notice that since $\log(1-2\lambda)$ decreases from $0$ to $-\infty$
as $\lambda$ increases from $0$ to $\frac{1}{2}$, $\lambda_0$
satisfies $\lambda_0\in (0,0.5)$. Since $\max_{0\leq j\leq m} S_j
\leq \max_{j\geq 0} S_j$, the bound in (\ref{umin1}) is also valid
for any {\it stopped} random walk.

\subsection{Upper bounds on $\gamma(\beta,k)$} \label{inc_gamma_ub}
For a scaled chi-square random variable $X$ with $k$ degrees of
freedom and mean $k\sigma^2$,
\[
\Pr(X\leq
\beta)=\gamma\left(\frac{\beta}{2\sigma^2},\frac{k}{2}\right)
\]
where $\gamma$ is known as the incomplete gamma function. From Chernoff bound, we have
\be \label{gamma_1}
\gamma\left(\frac{\beta}{2\sigma^2},\frac{k}{2}\right)=\Pr(-X\geq
-\beta)\leq \begin{cases} \left(\frac{\beta}{\sigma^2
k}\right)^{k/2} e^{(\frac{k}{2}-\frac{\beta}{2\sigma^2})} &
\frac{\beta}{2\sigma^2} <
\frac{k}{2} \\
1 &  \frac{\beta}{2\sigma^2} \geq \frac{k}{2}
\end{cases}
\ee A simpler, though looser, upper bound is given in
\cite{gautschi}: \be \label{gamma_2} \gamma(x,a)\leq (1-e^{-x})^a
\ee

%----------------------------
\subsection{Proof}
Let ${\mb x}_1^k$ be any path in the tree,
and $\displaystyle{h({\mb x}_1^k)=\sum_{i=1}^k w_i({\mb
x}_1^i)-bk}$, as in Section~\ref{stack_prp}. Let $f_M=\max_{1\leq i\leq m}
h(\hat{\xv}_1^i)$ be the maximum cost function along the transmitted path.
From Section~\ref{stack_prp}, Prop.~1, $f_M$ is not lesser than the maximum
of the cost functions along the path chosen by the stack decoder. From
Prop.~2, it is easy to see that any node ${\mb x}_1^k$ is generated, only
if the maximum of the cost functions along the path
${\mb x}_1^k$ does not increase above $f_M$. 

Let ${\mathcal A}_{s,b}$ be the set of generated nodes. In the proof, we upper
bound the number of all the paths visited by the algorithm that
are different from the correct path, and then we add the
complexity of finding the correct path (i.e., $m$). Then,
${\mathcal A}_{s,b}$ is a subset of the set of nodes that satisfy
$f({\mb x}_1^k)<f_M$. Let $\Rm_{k,k}$ be the lower $k\times k$ part of the
$\Rm$ matrix, {\it i.e.,}
\[
\Rm_{k,k}=\left(\begin{array}{ccc} r_{k,k} & \hdots & r_{k,1} \\  & \ddots & \vdots \\
\mb{0} &  & r_{1,1} \end{array}\right)
\]
Then, we have,
\begin{eqnarray}
P({\mb x}_1^k\in{\mc A}_{s,b})&\leq& P(|{\mb R}_{k,k}\tilde{\mb
  x}_1^k+{\mb w}_{r+1}^{r+k}|^2-bk<f_M)\\
&=& P\left(\left|\bmx{\mb R}_{k,k} \\ \mb{0}\emx\tilde{\mb x}_1^k+{\mb
w}_{1}^{r+k}\right|^2-bk<f_M+|\mb{w}_1^r|^2\right)\\
\label{stack-comp:1} &=&P\left(\left|\bmx {\mb R}_{k,k} \\ \mb{0}\emx\tilde{\mb x}_1^k\right|^2+|{\mb
    w}_{1}^{r+k}|^2+2({\mb w}_{1}^{r+k})^\transp \bmx {\mb R}_{k,k} \\ \mb{0}
\emx \tilde{\mb x}_1^k-bk<f_M+|\mb{w}_1^r|^2\right)
\end{eqnarray}
where $r=n-m$ is the excess degrees of freedom in the V-BLAST
system. From \cite{hassibi}, for each $k\leq m$, one can find an
$(r+k)\times k$ matrix $\bar{\mb H}_{r+k,k}$ that has the same distribution
as the lower $(r+k)\times k$ part of $\mb{H}$, and an $(r+k)\times(r+k)$
unitary matrix $\Thetam^{(r+k)}$ whose distribution is independent of $\Rm_{k,k}$, such that $\bar{\mb
  H}_{r+k,k}=\mb{\Theta}^{(r+k)}\begin{pmatrix}\Rm_{k,k} \\ \mb{0}
\end{pmatrix} $.

%  H}_{r+k,k}=\b{\Theta}^{(r+k)}\left(\begin{array}{c}{\mb R}_{k,k} \\
%    \mb{0}
%\end{array}\right)$.

Let $\bar{\mb
z}_1^{r+k}=\mb{\Theta}^{(r+k)}\mb{w}_1^{r+k}$.
The bound in (\ref{stack-comp:1}) can now be rewritten as
\begin{eqnarray}
\label{stack-comp:2} P({\mb x}_1^k\in{\mc
A}_{s,b})&\leq&P(|\bar{\mb H}_{r+k,k}\tilde{\mb
x}_1^k|^2+2(\bar{\mb z}_{1}^{r+k})^\transp\bar{\mb
H}_{r+k,k}\tilde{\mb x}_1^k<f_M+bk-|\mb{w}_{r+1}^{r+k}|^2)
\end{eqnarray}
In (\ref{stack-comp:2}), $f_M+bk-|\mb{w}_{r+1}^{r+k}|^2$ can be
bounded as
\begin{eqnarray}
f_M+bk-|{\mb w}_{r+1}^{r+k}|^2 &=& \max\{bk-|{\mb
w}_{r+1}^{r+k}|^2,b(k-1)-|{\mb w}_{r+2}^{r+k}|^2, \hdots, f'_M\}\\
&\leq& \max\{bk,f'_M\}
\end{eqnarray}
where $f'_M=\max\left\{0,|{\mb w}_{r+k+1}^{r+k+1}|^2-b,
\hdots,|{\mb w}_{r+k+1}^n|^2-b(m-k)\right\}$. Let
$\beta=\max\{bk,f'_M\}$. (\ref{stack-comp:2}) can now be rewritten
as
\begin{equation} \label{stack-comp:3}
P({\mb x}_1^k\in{\mc A}_{s,b})\leq  P(|\bar{\mb
H}_{r+k,k}\tilde{\mb x}_1^k|^2+2(\bar{\mb
z}_{1}^{r+k})^\transp\bar{\mb H}_{r+k,k}\tilde{\mb x}_1^k<\beta)
\end{equation}
Using Chernoff bound, (\ref{stack-comp:3}) can be written as
\begin{equation} \label{stack_comp:4}
P({\mb x}_1^k\in{\mc B}_{s,b}|q_k, \beta)\leq
\begin{cases}
e^{-(q_k-\beta)^2/(8q_k)}, & q_k > \beta \\
1, &  q_k < \beta \\
\end{cases}
\end{equation}
where $q_k=|\bar{\mb H}_{r+k,k}\tilde{\mb x}_1^k|^2$. Let
$\eta=\frac{\rho}{m}$. Then, in (\ref{stack_comp:4}),
$\frac{1}{\eta|\tilde{\mb x}_1^k|^2}{q_k}$ is a chi-square random
variable with $(r+k)$ degrees of freedom. Since the three random
variables, $\bar{\mb  H}_{r+k,k}\tilde{\mb
  x}_1^k$, ${\mb z}_1^k$ and $\beta$ are independent, averaging over
$q_k$ and $\beta$ gives
\begin{eqnarray}
P({\mb x}_1^k\in{\mc A}_{s,b}) &\leq&
E_{\beta}\left(\int_{0}^{\beta} f_{q_k}(q_k) dq_k +
\int_{\beta}^{\infty}e^{-(q_k-\beta)^2/(8q_k)}  f_{q_k}(q_k) dq_k
\right) \\
\label{stack_comp:5} &\leq& P(\beta=bk) \left(\int_{0}^{bk}
f_{q_k}(q_k) dq_k + \int_{bk}^{\infty} e^{-(q_k-bk)^2/(8q_k)}
f_{q_k}(q_k) dq_k \right) + P(\beta>bk)
\end{eqnarray}
In (\ref{stack_comp:5}), $P(\beta>bk)=P(f'_M>bk)\leq
e^{-bk\lambda_0}$ for $b>1$ (see Section~\ref{wald_ineq}). Note
that one requires $b>1$ because the distribution of the maximum of
the cost functions along the transmitted path will depend on $m$
otherwise. \footnote{Later, we will require a stronger condition on $b$ to
guarantee the convergence of the sums in (\ref{stack_comp:8}).} The bound
in (\ref{stack_comp:5}) amounts to counting all the nodes $\xv_1^k$ in the
search space when $\beta>bk$. Since $P(\beta>bk)$ decreases sufficiently
fast as $k$ increases, this upper bound is still tight for our purposes.
Now, \eqref{stack_comp:5} can be further simplified as,
\begin{eqnarray}
\label{stack_comp:6} P({\mb x}_1^k\in{\mc A}_{s,b})&\leq&
\int_{0}^{bk} f_{q_k}(q_k) dq_k  + \int_{bk}^{\infty}
e^{-(q_k-bk)^2/(8q_k)} f_{q_k}(q_k) dq_k +
e^{-bk\lambda_0} \\
&\leq&  \int_{0}^{bk} f_{q_k}(q_k) dq_k  + \int_{0}^{\infty}
e^{-(q_k-bk)^2/(8q_k)} f_{q_k}(q_k) dq_k +
e^{-bk\lambda_0} \\
&\leq& \gamma\left(\frac{bk}{2\eta |\tilde{\mb
      x}_1^k|^2},\frac{r+k}{2}\right) +
e^{bk/4}\int_{0}^{\infty} e^{-q_k/8}f_{q_k}(q_k) dq_k +
e^{-bk\lambda_0} \\
\label{stack_comp:7a} &\leq&
\gamma\left(\frac{bk}{2\eta},\frac{r+k}{2}\right) +
\frac{e^{bk/4}}{\left(1+\frac{\eta}{4}\right)^{(r+k)/2}} +
e^{-bk\lambda_0}
\end{eqnarray}
with $\gamma(\cdot,\cdot)$ as the incomplete gamma function.
Assuming $r\geq 0$, (\ref{stack_comp:7a}) can be bounded as
\begin{eqnarray}
\label{stack_comp:7} P({\mb x}_1^k\in{\mc A}_{s,b})&\leq&
\left(\frac{b}{\eta}e^{1-\frac{b}{\eta}}\right)^{\frac{k}{2}} +
\left(\frac{e^{b/2}}{(1+\frac{\eta}{4})}\right)^{\frac{k}{2}} +
e^{-bk\lambda_0}
\end{eqnarray}
for $\eta>b$. The inequality in (\ref{stack_comp:7}) follows from
an upper bound on the incomplete gamma function (see
Section~\ref{inc_gamma_ub}). For a node ${\mb x}_1^k$, let $G({\mb
x}_1^k)=1$ if the node is generated and $0$ otherwise. Then, the
expected number of nodes generated by the algorithm ({\it i.e.,}
complexity) is $\displaystyle{\sum_{k=1}^{m}
  \sum_{{\mb
    x}_1^k} E[G({\mb x}_1^k)]}$, where the
expectation is over all channel realizations. Let $C_m$ be the
expected complexity per dimension. Then, assuming a bounded $r$,
$C_m$ is written as \footnote{The first term in the RHS
  of \eqref{boundCm} comes from counting the complexity of the
finding correct path, i.e., $\tilde{\mb x}=\mb{0}$.}
\begin{eqnarray}\label{boundCm}
mC_m\leq m+\sum_{k=1}^{m} \sum_{{\mb x}_1^k} \left(
  \left(\frac{b}{\eta}e^{1-\frac{b}{\eta}}\right)^{\frac{k}{2}} +
\left(\frac{e^{b/2}}{(1+\frac{\eta}{4})}\right)^{\frac{k}{2}} +
e^{-bk\lambda_0} \right).
\end{eqnarray}
The complexity per dimension, $C_m$, can now be upper bounded as
\begin{eqnarray}
C_m&\leq& 1+\frac{1}{m}\sum_{k=1}^{m} \sum_{{\mb x}_1^k} \left(
  \left(\frac{b}{\eta}e^{1-\frac{b}{\eta}}\right)^{\frac{k}{2}} +
\left(\frac{e^{b/2}}{(1+\frac{\eta}{4})}\right)^{\frac{k}{2}} +
e^{-bk\lambda_0} \right) \\
\label{stack_comp:8a} &\leq& 1 +\frac{1}{m}\sum_{k=1}^{m} Q^k
\left(
  \left(\frac{b}{\eta}e^{1-\frac{b}{\eta}}\right)^{\frac{k}{2}} +
\left(\frac{e^{b/2}}{(1+\frac{\eta}{4})}\right)^{\frac{k}{2}} +
e^{-bk\lambda_0} \right) \\
\label{stack_comp:8} &\leq& 1 +
\frac{1}{m}\left(\frac{1}{1-Q^2\frac{b}{\eta}e^{1-\frac{b}{\eta}}}+\frac{1}{1-Q^2\frac{e^{b/2}}{(1+\frac{\eta}{4})}}+\frac{1}{1-Qe^{-b\lambda_0}}\right),
\end{eqnarray}
when $b$ and $\eta$ are sufficiently large, so that all the three
sums converge. The inequality in (\ref{stack_comp:8a}) is true,
since the number of nodes at level $k$ is $Q^k$.  Since the terms
inside the parenthesis in (\ref{stack_comp:8}) are all independent
of $m$, the number of nodes visited by the stack algorithm scales
at most linearly, when $\eta>\eta_0$, where $\eta_0$ is the
minimum $\frac{\rho}{m}$ ratio required for convergence of the
sums in (\ref{stack_comp:8}). \hfill $\Box$

%---------------------------------------------------------------------------------

\begin{table}[!h]\label{table1}
\caption{Complexity Ratio of the proposed algorithm
  for Rate-$3$ TAST constellation over Rate-$1$ TAST constellation in
a $3\times 1$ MIMO System}
\begin{center}
\(\begin{array}{|c||c|c|c|c|c|}
\hline & & & & &  \\
\mbox{SNR (dB)} & 22 &  24 &  26 &  28 & 30
\\ \hline \gamma & 41 &  31 &
  23 &  16 &  12 \\ \hline \end{array}\)
\end{center}
\end{table}

%----------------------------------------------------------------------------------

%---------------------------------------------------------------------------------

%----------------------------------------------------------------------------------

%--------------------------------------------------------------------------------

%------------------------------------------------------------------------------

\bibliographystyle{abbrv}
\bibliography{seq_4}

\begin{thebibliography}{10}

\bibitem{agrell}
E.~Agrell, T.~Eriksson, A.~Vardy, and K.~Zeger.
\newblock Closest point search in lattices.
\newblock {\em {IEEE Transactions on Information Theory}}, 48(8):2201--2214,
  August 2002.

\bibitem{anderson}
J.~B. Anderson and S.~Mohan.
\newblock Sequential coding algorithms: A survey and cost analysis.
\newblock {\em {IEEE Trans. Comm.}}, 32:169--176, Feb. 1984.

\bibitem{babai}
L.~Babai.
\newblock On lovasz lattice reduction and the nearest lattice point problem.
\newblock {\em {Combinatorica}}, 6(1):1--13, 1986.

\bibitem{hagenauer}
S.~Baro, J.~Hagenauer, and M.~Witzke.
\newblock Iterative detection of {MIMO} transmission using a list-sequential
  ({LISS}) detector.
\newblock In {\em Proc. IEEE Int. Conf. Communications}, pages 2653--2657, May
  2003.

\bibitem{golden-code}
J.-C. Belfiore, G.~Rekaya, and E.~Viterbo.
\newblock The golden code: A$2\times 2$ full-rate space-time code with
  nonvanishing determinants.
\newblock {\em {IEEE Transactions on Information Theory}}, 51(4):1432 -- 1436,
  Apr. 2005.

\bibitem{fast-vblast}
J.~Benesty, Y.~A. Huang, and J.~Chen.
\newblock A fast recursive algorithm for optimum sequential signal detection in
  a blast system.
\newblock {\em {IEEE Trans. Signal Processing}}, 51:1722--1731, July 2003.

\bibitem{caire_psp}
G.~Caire and G.~Colavolpe.
\newblock On low complexity space-time coding for quasi-static channels.
\newblock {\em {IEEE Trans. Info. Theory}}, 49(6):1400--1416, June 2003.

\bibitem{forney-cioffi-mmsegdfe}
J.~M. Cioffi, G.~P. Dudevoir, M.~V. Eyuboglu, and G.~D. {Forney Jr.}
\newblock {MMSE} decision-feedback equalizers and coding. {I}. equalization
  results.
\newblock {\em {IEEE Transactions on Communications}}, 43(10):2582 -- 2594,
  Oct. 1995.

\bibitem{cohen}
H.~Cohen.
\newblock {\em A Course in Computational Algebraic Number Theory}.
\newblock Springer-Verlag, 1995.

\bibitem{splug}
J.~H. Conway and N.~J.~A. Sloane.
\newblock {\em Sphere Packings, Lattices, and Groups, 3rd ed.}
\newblock Springer-Verlag New York, 1999.

\bibitem{gsd}
M.~O. Damen, K.~Abed-Meraim, and J.-C. Belfiore.
\newblock Generalized sphere decoder for asymmetrical space-time communication
  architecture.
\newblock {\em { Electron. Lett}}, 36:166, Jan. 2000.

\bibitem{damen0}
M.~O. Damen, A.~Chkeif, and J.-C. Belfiore.
\newblock Lattice codes decoder for space-time codes.
\newblock {\em {IEEE Commun. Lett.}}, 4:161--163, May 2000.

\bibitem{tast-constellations}
M.~O. Damen, H.~{El Gamal}, and N.~C. Beaulieu.
\newblock Linear threaded algebraic space-time constellations.
\newblock {\em {IEEE Transactions on Information Theory}}, 49(10):2372--2388,
  Oct. 2003.

\bibitem{damen1}
M.~O. Damen, H.~{El~Gamal}, and G.~Caire.
\newblock On maximum-likelihood detection and the search for the closest
  lattice point.
\newblock {\em {IEEE Transactions on Information Theory}}, 49:2389--2401, Oct.
  2003.

\bibitem{damen2}
M.~O. Damen, H.~{El Gamal}, and G.~Caire.
\newblock {MMSE-GDFE} lattice decoding for under-determined linear channels.
\newblock In {\em 38th Annual Conf. on Inform. Sciences and Systems}, March
  2004.

\bibitem{Mahesh}
P.~Dayal and M.~K. Varanasi.
\newblock A fast generalized sphere decoder for optimum decoding of
  under-determined {MIMO} systems.
\newblock In {\em Proc. 41th Annual Allerton Conf. on Comm. Control, and
  Comput., Monticello, IL}, Oct. 2003.

\bibitem{isi}
A.~{Duel-Hallen} and C.~Heegard.
\newblock Delayed decision-feedback sequence estimation.
\newblock {\em {IEEE Transactions on Communications}}, 37:428--436, May 1989.

\bibitem{hesham1}
H.~{El Gamal}, G.~Caire, and M.~O. Damen.
\newblock Lattice coding and decoding achieve the optimal
  diversity-multiplexing tradeoff of {MIMO} channels.
\newblock {\em {IEEE Transactions on Information Theory}}, 50(6):968--985, June
  2004.

\bibitem{tast}
H.~{El Gamal} and M.~O. Damen.
\newblock Universal space-time coding.
\newblock {\em {IEEE Transactions on Information Theory}}, 49:1097--1119, May
  2003.

\bibitem{erez-zamir-litsyn}
U.~Erez, S.~Litsyn, and R.~Zamir.
\newblock Lattices which are good for (almost) everything.
\newblock In {\em Proc. IEEE Information Theory Workshop, 2003}, pages
  271--274, Apr. 2003.

\bibitem{erez-zamir}
U.~Erez and R.~Zamir.
\newblock Achieving $\frac{1}{2} \log(1+{SNR})$ on the {AWGN} channel with
  lattice encoding and decoding.
\newblock {\em {IEEE Transactions on Information Theory}}, 50(10):2293 -- 2314,
  Oct. 2004.

\bibitem{fano}
R.~M. Fano.
\newblock A heuristic discussion of probabilistic decoding.
\newblock {\em {IEEE transactions on Information Theory}}, 9(2):64--74, Apr.
  1963.

\bibitem{pohst}
U.~Fincke and M.~Pohst.
\newblock Improved methods for calculating vectors of short length in a
  lattice, including a complexity analysis.
\newblock {\em {Math. of Comput.}}, 44:463--471, Apr. 1985.

\bibitem{forney-coset1}
G.~D. {Forney Jr.}
\newblock Coset codes. {I}. introduction and geometrical classification.
\newblock {\em {IEEE Transactions on Information Theory}}, 34(5):1123 -- 1151,
  Sept. 1988.

\bibitem{forney-coset2}
G.~D. {Forney Jr.}
\newblock Coset codes. {II}. binary lattices and related codes.
\newblock {\em {IEEE Transactions on Information Theory}}, 34(5):1152 -- 1187,
  Sept. 1988.

\bibitem{foschini}
G.~Foschini.
\newblock Layered space-time architecture for wireless communication in a
  fading environment using multi-element antennas.
\newblock {\em {Bell labs Tech J.}}, 1(2):41--59, 1996.

\bibitem{vblast}
J.~Foschini, G.~Golden, R.~Valenzuela, and P.~Wolniansky.
\newblock Simplified processing for high spectral efficiency wireless
  communication employing multi-element arrays.
\newblock {\em { IEEE J. Select. Areas Commun.}}, 17(11):1841--1852, Nov. 1999.

\bibitem{gautschi}
W.~Gautschi.
\newblock The incomplete gamma functions since {Tricomi}.
\newblock In {\em Tricomi's ideas and contemporary applied mathematics, Atti
  Convegni Lincei, Rome}, pages 203--237, 1998.

\bibitem{geist}
J.~Geist.
\newblock Search properties of some sequential decoding algorithms.
\newblock {\em {IEEE Transactions on Information Theory}}, 19(4):519--526, July
  1973.

\bibitem{radhika1}
R.~Gowaikar and B.~Hassibi.
\newblock Efficient statistical pruning for maximum likelihood decoding.
\newblock In {\em Proceedings of the 2003 IEEE International Conference on
  Acoustics, Speech and Signal Processing}, pages V--49--52, April 2003.

\bibitem{RHHG}
A.~R. {Hammons Jr} and H.~{El Gamal}.
\newblock On the theory of space-time codes for {PSK} modulation.
\newblock {\em {IEEE Transactions on Information Theory}}, 46(2):524 -- 542,
  March 2000.

\bibitem{hassibi-squareroot}
B.~Hassibi.
\newblock An efficient square-root algorithm for {BLAST}.
\newblock In {\em Proceedings of the 2000 IEEE International Conference on
  Acoustics, Speech and Signal Processing}, pages II737 -- II740, June 2000.

\bibitem{LD}
B.~Hassibi and B.~Hochwald.
\newblock High-rate codes that are linear in space and time.
\newblock {\em {IEEE Transactions on Information Theory}}, 48(7):1804 -- 1824,
  July 2002.

\bibitem{hassibi}
B.~Hassibi and H.~Vikalo.
\newblock On sphere decoding algorithm. {I}. expected complexity.
\newblock {\em {Submitted to IEEE Transactions on Signal Processing}}.

\bibitem{hoch-ten}
B.~Hochwald and S.~ten Brink.
\newblock Achieving near-capacity on a multiple-antenna channel.
\newblock {\em {IEEE Transactions on Communications}}, 51(3):389 -- 399, March
  2003.

\bibitem{otter}
J.~Jalden and B.~Ottersten.
\newblock On the complexity of sphere decoding in digital communications.
\newblock {\em {IEEE Trans. Signal Proc.}}, 53(4):1474 -- 1484, Apr. 2005.

\bibitem{johann}
R.~Johannesson and K.~Zigangirov.
\newblock {\em Fundamentals of convolutional coding}.
\newblock Wiley-IEEE press, 1999.

\bibitem{kokkonen}
M.~Kokkonen and K.~Kalliojarvi.
\newblock Soft-decision decoding of binary linear codes using the
  $t$-algorithm.
\newblock In {\em Proc. IEEE $8$th intl. symp. on PIMRC}, pages 1181--1185,
  Sep. 1997.

\bibitem{LLL}
A.~K. Lenstra, A.~W. {Lenstra Jr.}, and L.~Lovasz.
\newblock On factoring polynomials with rational coefficients.
\newblock {\em {Math. Annalen.}}, 261:515--534, 1982.

\bibitem{LFT}
Y.~Liu, M.~Fitz, and O.~Takeshita.
\newblock A rank criterion for qam space-time codes.
\newblock {\em {IEEE Trans. Info. Theory}}, 48(12):3062--3079, Dec. 2002.

\bibitem{LK}
H.-F. Lu and P.~Kumar.
\newblock A unified construction of space-time codes with optimal
  rate-diversity tradeoff.
\newblock {\em {IEEE Transactions on Information Theory}}, 51(5):1709 -- 1730,
  May 2005.

\bibitem{schn-euch}
C.~P. Schnorr and M.~Euchner.
\newblock Lattice basis reduction: Improved practical algorithms and solving
  subset sum problems.
\newblock {\em {Math. Programming}}, 66:181--191, 1994.

\bibitem{division-algebra}
B.~A. Sethuraman, B.~S. Rajan, and V.~Shashidhar.
\newblock Full-diversity, high-rate space-time block codes from division
  algebras.
\newblock {\em {IEEE Transactions on Information Theory}}, 49(10):2596 -- 2616,
  Oct. 2003.

\bibitem{nee}
R.~van Nee, A.~van Zelst, and G.~Awater.
\newblock Maximum likelihood decoding in a space division multiplexing system.
\newblock In {\em Vehicular Technology Conference}, pages 6--10, May 2000.

\bibitem{vit-bou}
E.~Viterbo and J.~Boutros.
\newblock A universal lattice code decoder for fading channels.
\newblock {\em {IEEE Transactions on Information Theory}}, 45(5):1639--1642,
  1999.

\bibitem{windpass}
C.~Windpassinger and R.~Fischer.
\newblock Low-complexity near-maximum-likelihood detection and precoding for
  {MIMO} systems using lattice reduction.
\newblock In {\em Proc. IEEE Inform. Theory Workshop, Paris, France,}, Mar.
  2003.

\bibitem{wozencraft}
J.~M. Wozencraft and B.~Reiffen.
\newblock {\em Sequential decoding}.
\newblock MIT press and Wiley, 1961.

\bibitem{Xu}
W.~Xu, Y.~Wang, Z.~Zhou, and J.~Wang;.
\newblock A computationally efficient exact {ML} sphere decoder.
\newblock In {\em IEEE Global Telecommunications Conference, 2004.}, pages 2594
  -- 2598, Nov 2004.

\bibitem{wornell}
H.~Yao and G.~Wornell.
\newblock Lattice-reduction-aided detectors for {MIMO} communication systems.
\newblock In {\em Proc. IEEE Global Conf. on Commun., Taipei, Taiwan}, Nov.
  2002.

\end{thebibliography}

\end{document}